\documentclass[12pt]{article}
\pdfoutput=1
\usepackage[totalwidth=480pt,totalheight=640pt]{geometry}
\usepackage[centertags]{amsmath}
\usepackage[square, comma, sort&compress,numbers]{natbib}
\usepackage{array,multirow}
\usepackage{slashed}
\usepackage{amssymb}
\usepackage{graphicx}
\usepackage{color}
\usepackage{setspace}

\usepackage{epsfig}

\def\Or[#1]{{\text{O}}\left({#1}\right)}
\def\dotl[#1,#2]{\left\langle #1, #2 \right\rangle}
\def\dotlb[#1,#2]{[ #1, #2 ]}
\def\dotp[#1,#2]{(#1) \cdot (#2)}
\def\aff[#1,#2]{\hat{#1}(#2)}
\def\n4sym{{\cal N}=4 SYM}
\def\>{\rangle}
\def\<{\langle}
\def\weight[#1,#2,#3]{\{(#1),#2,#3\}}
\def\ads[#1]{$\text{AdS}_{#1}$}

\newcommand{\ba}{\begin{eqnarray}}
\newcommand{\ea}{\end{eqnarray}}
\newcommand{\be}{\begin{eqnarray}}
\newcommand{\ee}{\end{eqnarray}}

\newcommand{\CC}{{\cal C}}

\newcommand{\CO}{{\cal O}}

\newcommand{\nn}{\nonumber}

\usepackage{hyperref}

\begin{document}

\begin{titlepage}

\begin{center}
\vspace{1cm}

%Effective Field Theory for Holographic Duality with Broken Conformal Symmetry

{\Large \bf Decoupling of High Dimension Operators from \\ the Low Energy Sector in Holographic Models}
\vspace{0.8cm}

\small
\bf{A. Liam Fitzpatrick$^1$,  Jared Kaplan$^{1,2}$, Emanuel Katz$^3$, Lisa Randall$^4$}
\normalsize

\vspace{.5cm}

{\it $^1$ Stanford Institute for Theoretical Physics, Stanford University, Stanford, CA 94305}\\
{\it $^2$ Department of Physics and Astronomy, Johns Hopkins University, Baltimore, MD 21218} \\
{\it $^3$ Physics Department, Boston University, Boston, MA 02215, USA} \\
{\it $^4$ Department of Physics, Harvard University, Cambridge, MA 02138} \\

\end{center}

\vspace{1cm}

\begin{abstract}

We study the decoupling of high dimension operators from the the description of the low-energy spectrum in theories where conformal symmetry is broken by a single scale, which we refer to as `broken CFTs'.  Holographic duality
suggests that this decoupling occurs in generic backgrounds.  
We show how the decoupling of high mass states in the $(d+1)$-dimensional bulk relates to the decoupling of high energy states in the $d$-dimensional broken CFT.   In other words, we explain why both high dimension operators and high mass states in the CFT  decouple from the low-energy physics of the mesons and glueballs.  In many cases, the decoupling can occur exponentially fast in the dimension of the operator.   
Holography motivates a new kind of form factor proportional to the two point function between broken CFT operators with  very different scaling dimensions.   This new notion of decoupling can provide a systematic justification for holographic descriptions of QCD and condensed matter systems with only light degrees of freedom in the bulk. %  {\blue Mention reduction of dimensions can lead to NEC violation?}

\end{abstract}

\bigskip

\end{titlepage}

\section{Introduction}

Our understanding of modern physics is largely based on Effective Field Theory (EFT), which allows us to obtain universal predictions about long-distances while decoupling the short-distance details.  The AdS/CFT correspondence \cite{Maldacena:1997re, Witten, Gubser:1998bc} and the development of Randall-Sundrum models \cite{Randall:1999ee, Randall:1999vf}, AdS/QCD \cite{Karch:2002sh, Erlich:2005qh, Karch:2006pv}, and even AdS/CMT \cite{Hartnoll:2008vx, Balasubramanian:2008dm, Son:2008ye, Liu:2009dm} has led to the study of EFT in a qualitatively new context.  Although it was implicit in many earlier works, the systematic study of EFT in AdS/CFT has been a relatively recent development, perhaps beginning with \cite{JP, Katz} and continuing with many further studies \cite{Fitzpatrick:2011hh, NaturalLanguage, ElShowk:2011ag, AdSfromCFT}.  

In this work we will study field theories such as QCD that are approximately conformal at short distances, but that break conformal invariance at long-distances and generically have a mass gap.  One can naturally describe these `Broken CFTs' at a long distances with an EFT in $d$-dimensions, where we will often refer to the familiar case $d=4$.  In the case of QCD, this would simply be the EFT of pions, and perhaps also $\rho$ mesons, nucleons, etc.  However, in light of AdS/CFT, it becomes natural to consider an alternative description in terms of a  $(d+1)$-dimensional warped space approximating AdS.  The description in the warped space can also be an EFT with only a finite number of light bulk fields.   If the CFT has certain special properties \cite{JP, Heemskerk:2010ty,  ElShowk:2011ag, Sundrum:2011ic, AdSfromCFT} then this can be a very good description.  

The challenge in making this correspondence is that invariant masses in the warped bulk spacetime are not directly associated with $d$-dimensional energies; instead they are related to the dimensions of operators in the broken CFT \cite{Fubini:1972mf, Katz}. More formally, this connection is explained by the fact that eigenstates of the AdS Laplacian map directly to eigenstates of the Conformal Casimir in the CFT, so that
\be
\nabla^2_{AdS} \longleftrightarrow \mathcal{C}_2 = D^2 + P_\mu K^\mu + K_\mu P^\mu + M_{\mu \nu}^2
\ee
with eigenvalues $m^2 R_{AdS}^2 = \Delta(\Delta-4)$ for scalar operators.\footnote{For CFT primaries of dimension $\Delta$ and spin $\ell$ the conformal Casimir has eigenvalues $\Delta(\Delta-d) + \ell(\ell + d -2)$.}  Consequently, there are two distinct notions of EFT, the traditional kind based on the $d$-dimensional masses of particle resonances in the Broken CFT, and a new kind, based on the $(d+1)$-dimensional masses of bulk fields.  Because bulk masses are dual to the dimensions of operators in the CFT,  integrating out high mass bulk states leads to an Effective CFT \cite{Katz} with a cutoff in scaling dimensions.   The relation between the $d+1$-dimensional EFT or `ECFT'   \cite{Katz}  and the more standard $d$-dimensional EFT is non-trivial.  It is not even clear that the decoupling of the ``short-distance'' physics in one theory will be compatible with decoupling in the other.  The purpose of this work will be to study the relationship between these two quite different effective field theory descriptions.  

Although we are motivated by AdS/CFT, many of our central questions can be phrased purely in the language of the broken CFT.    For example, consider two CFT operators $\CO_1$ and $\CO_2$ of dimension $\Delta_1 \neq \Delta_2$.  When conformal symmetry is unbroken the 2-pt function of $\CO_1$ with $\CO_2$ vanishes, but in the presence of a mass gap we expect\footnote{We discuss operator normalizations, which are crucial for defining the magnitude of this correlator, in appendix \ref{app:Normalizations}.} 
\be
\langle \CO_2(r) \CO_1(0) \rangle \approx f(\Delta_1, \Delta_2) \frac{e^{-m r}}{r^{d-2}}
\ee
in the large $r$ limit, where $m$ is the mass of the lightest particle created by $\CO_1$ and $\CO_2$.  We can now ask how $f(\Delta_1, \Delta_2)$ behaves in the limit that $\Delta_2 \gg \Delta_1$, in particular when $\Delta_1 \sim$ few, so we simply have $f_{\Delta_1}(\Delta_2)$.  One of our main goals will be to show that we must have $f_{\Delta_1}(\Delta) \to 0$ as $\Delta \to \infty$ and to understand the physics of the rate.  We will see that we may have a power-law dependence such as $f_{\Delta_1}(\Delta) \propto 1 / \Delta^2$ in some cases, for example in RS-type models.  However, in many situations we actually find an exponential dependence
\be
\label{eq:TwoPointFormFactor}
f_{\Delta_1}(\Delta) \sim \exp \left[ - \lambda \Delta^p \right]
\ee
where the rate depends on the density of states in the sector created by $\CO_1$ and $\CO_2$, and $\lambda$ is some constant that may depend on $\CO_1$.  In the case of linearly confining theories such as `soft-wall' AdS/QCD \cite{Karch:2006pv}, we predict that $p=1$ from the arguments of section \ref{sec:GeneralDecoupling}.  When $f \to 0$ quickly at large $\Delta$, we have a rapid decoupling of the high dimension operators from the low-dimension {\it and} low-mass spectrum of the broken CFT.  The latter follows because the probability for a high dimension operator to create very light particles is proportional to $|f_{\Delta_1}(\Delta)|^2$.  This makes the $d$ and $d+1$ dimensional notions of decoupling compatible. 

We will argue that in bulk models that are asymptotically AdS in the UV, large dimension CFT operators naturally decouple from the interactions of the light particles, which we refer to as `mesons and glueballs'.  In the bulk theory, there are two distinct mechanisms behind this decoupling:
\begin{itemize}
\item Higher dimension operators in the broken CFT can create mesons with larger masses, so $d$-dimensional decoupling follows from $(d+1)$-dimensional decoupling.
\item Bulk modes associated with operators of different dimensions can be localized in different regions.  Bulk locality then leads higher dimension operators to decouple exponentially as a consequence of tiny wave-function overlaps.
\end{itemize}
We will see that on the one hand, the first mechanism occurs in ``hard wall'' RS-type models, where space-time ends at some point in the bulk. 
The second mechanism tends to dominate in generic ``soft wall'' models, where the bulk geometry cuts off more smoothly.  Our results explain the decoupling of high-dimension operators in explicitly solvable models such as 2-d QCD \cite{tHooft:1974hx, Katz:2007br}.

Our analysis will mostly be at the level of the quadratic bulk actions that determine CFT operator dimensions, meson and glueball masses, and bulk wavefunctions.   These data can be determined from a one-dimensional Schrodinger equation for the bulk modes.  Our results will be sufficient to demonstrate the decoupling of high dimension operators in large $N$ type broken CFTs, for it will imply that the interactions of high-dimension operators must be suppressed by a combination of large energy denominators and small couplings from suppressed bulk wavefunction overlaps.  We will leave a more detailed study of the meson and glueball interactions for future work.

The outline of this paper is as follows.  In section \ref{sec:RSModelReview} we will review a number of results about hard-wall RS-type models and holography, emphasizing the role of effective field theory in the bulk.  In section \ref{sec:HardWall} we will give a detailed analysis of the expectations from naturalness for meson and glueballs in hard-wall models.  We move on to study general models in section \ref{sec:GeneralWall}, beginning by motivating more general warped metrics with dilaton profiles.  We explain the mode decomposition in section \ref{sec:Wavefunctions}.  Instead of attempting to solve the Einstein's equations, we study and then utilize model-independent constraints from the Null Energy Condition on the metric, which are dual to the $a$-theorem \cite{Freedman:1999gp,Girardello:1998pd,Myers:2012ed}.  We conclude the section with a few simple examples.   We then give a general argument for decoupling in section \ref{sec:GeneralDecoupling}.  Since this argument shows that operators with very different dimensions typically create mesons localized in different regions of the bulk, in section \ref{sec:Sizes} we explain why this does not lead to very different physical meson sizes in the $d$-dimensional theory.  We discuss our results in section \ref{sec:Discussion} and emphasize the importance of further investigating these phenomenon directly in the CFT.  In appendix \ref{sec:JustifyingQuadratic} we justify an approximation used in section \ref{sec:GeneralDecoupling} and give a brief demonstration of how the form factor in equation (\ref{eq:TwoPointFormFactor}) relates to the density of states in a simple class of models.  In appendix \ref{app:Normalizations} we explain CFT operators normalizations.

\section{Review and Observations About RS Models}
\label{sec:RSModelReview}

We will begin by making some simple observations about the relationship between $d+1$ and $d$-dimensional effective field theory and naturalness in RS-type models \cite{Randall:1999ee, Randall:1999vf}.  RS models have been extremely well-studied and reviewed, see e.g. \cite{Gherghetta:2006ha}, and most of the points we will make here are known to experts; we assemble them as an introduction to a more general story.  For notational convenience and familiarity we will usually take $d=4$, although the points we make will not be specific to this case. 

RS models live in the Poincar\'e patch of AdS space, with metric
\be
ds^2 = \frac{1}{(kz)^2} \left( \eta^{\mu \nu} dx_{\mu} dx_{\nu} - dz^2  \right),
\ee 
where $k=R_{\rm AdS}^{-1}$ is the AdS curvature scale. The AdS slice ends at an `IR brane' at $z=z_{IR}$ in the bulk, and this brane is the source of conformal symmetry breaking.  Because the AdS space abruptly ends at $z_{IR}$, RS models are also referred to as `hard wall' models.  We will only be studying the dynamics of the conformal sector by itself, so we will not include a UV brane.  In RS models we have a `radion' or 4-d dilaton mode corresponding, in effect, to fluctuations of $z_{IR}$.  The radion is a single 4-d degree of freedom that arises when conformal symmetry is broken.  If we have spontaneous breaking of conformal invariance the radion would be a massless goldstone boson, whereas a massive radion would indicate the explicit breaking of conformal invariance.  In generic non-supersymmetric broken CFTs we expect to have explicit breaking with a massive radion, but at large $N$ the radion might still be identifiable, as it will have a small width.

Most of the degrees of freedom in an RS model will be fields that live in the bulk of the AdS space, corresponding to operators in the CFT that would fill out full irreducible representations of the conformal group in the absence of conformal symmetry breaking.  These fields will generically have both a bulk action and a boundary action localized on the IR brane at $z_{IR}$.  Given a cutoff $\Lambda_{5}$ for the 5-d EFT in the bulk, we expect that the bulk and boundary actions will generically contain all operators consistent with the symmetries of the theory, with coefficients related to powers of $\Lambda_5$ with order one coefficients.  For example, for a scalar field $\Phi$ we would most naively expect
\be
\label{eq:EFTAction}
S &=& \int_{bulk} d^4 x dz \sqrt{-g} \left( \frac{1}{2} (\nabla_A \Phi )^2 - \frac{1}{2} c_2 \Lambda_5^2 \Phi^2 - c_3 \sqrt{\Lambda_5} \Phi^3 + c_4 \frac{1}{\Lambda_5} \Phi^4 + \cdots \right)  
\nonumber \\ && + \int_{z_{IR}} d^4 x \left(\frac{1}{2} a_{\perp}  \partial_z  ( \Phi^2) + \frac{1}{2} a_2 \Lambda_5 \Phi^2 + \frac{1}{2} a_{\parallel} \frac{1}{\Lambda_5} (\partial_\mu \Phi)^2 + \cdots \right)
\ee
The bulk mass-squared term $c_2 \Lambda_5^2$ in the above Lagrangian is related to the dimension $\Delta$ of dual CFT operator by the usual relation for scalars $c_2 \Lambda_5^2R_{\rm AdS}^2 = \Delta(\Delta-4)$. We have made the assumption that the potential on the IR brane does not have any linear `tadpole' terms.  This means that it is minimized with $\Phi = 0$, so that $\Phi$ will not acquire a non-trivial bulk profile breaking conformal symmetry away from the IR brane.  Alternatively, fields with $\Phi(z_{IR}) \neq 0$ but $\Phi(z) \to 0$ as $z \to 0$ can be used to produce a radion potential, as in the Goldberger-Wise mechanism \cite{GoldbergerWise}. 
In many constructions the dimensionless couplings $c_i$ and $a_i$ can be small because of symmetries, or due to an overall factor of $1/N$ in the CFT.  In AdS this can correspond to the difference between e.g. the string scale and the Planck scale.  

Note that the IR brane localized action on the second line of equation (\ref{eq:EFTAction}) contains precisely two marginal or relevant terms; although we have written the action in $d=4$ these two terms would be present in any number of dimensions.   Keeping only these terms and a contribution from the bulk action, one obtains one of the two boundary conditions:
\be
 \Phi(z_{IR}) = 0, && \qquad  \textrm{(Dirichlet)}, \nn\\
\left. (a_\perp - 1) \partial_z \Phi + a_2 \Lambda_5 \Phi \right|_{z_{IR}} = 0, && \qquad  \textrm{(Mixed)}.
\ee
In the limit that $a_\perp = 1$, both Dirichlet and Mixed boundary conditions set $\Phi=0$ at $z_{IR}$, because the boundary kinetic term cancels a contribution from the kinetic term in the bulk action, leaving only the boundary mass term.  If $a_2 = 0$, then we have the special case of a Neumann boundary condition that imposes $\partial_z  \Phi=0 $ at $z_{IR}$.  It is the combination of the bulk mass term $c_2$ with this boundary condition that determines the KK mode masses. The ultimate spectrum of 4d masses depends on both the dynamics of the purely conformal sector together with some details of the conformal symmetry breaking, reminiscent of the way the low-energy spectrum in theories with broken supersymmetry depends on the mechanism of supersymmetry breaking.  Our main focus will be the limit of large bulk masses when the conformal symmetry breaking dynamics are kept fixed.

As shown long ago in \cite{Georgi:2000ks, Cheng:2002iz}, unless we choose Dirichlet boundary conditions we do not have the option of ignoring the $a_i$ couplings.  Loop corrections in the bulk theory give rise to divergences that require counterterms for the $a_i$ couplings.  Even if we tune some of them to zero, the marginal couplings are regenerated by logarithmic running.

\section{Hard Wall Models and Effective Field Theory}
\label{sec:HardWall}

In the following two sections we will explain why decoupling in $(d+1)$-dimensional RS or `hard wall' models is compatible with decoupling in the $d$-dimensional EFT description of the low-mass `mesons' and `glueballs'.  In particular, we recover the standard result that states of large mass $M_{d+1}$ in the $(d+1)$-dimensional theory, which are dual to high dimension CFT operators, create states with $d$-dimensional masses $m_d \propto M_{d+1}$.    We treat both scalars and fermions in order to address the role of chiral symmetry, which protects fermion masses.

These results have implications for RS models \cite{Randall:1999ee} and hard-wall AdS/QCD \cite{Erlich:2005qh}.  For example, since the Higgs boson must be light compared to other resonances in RS, we should expect it to be described by a low-mass bulk field, which suggests that it will not be sharply localized on the IR brane.  A related phenomenon was recently seen in \cite{Delaunay:2012cz}.

\subsection{Scalars}

We will now show that in the absence of symmetry or tuning, the lightest KK mode from a heavy bulk field has a large $d$-dimensional mass.  This is the simplest example of decoupling. 
 To quantify this, we can consider how this $d$-dimensional mass changes as we vary the boundary conditions on the IR brane.  

Let us start with an easy special case in $d=4$ dimensions before treating the general problem.  We will look at how perturbations to the marginal and relevant quadratic terms on the IR brane give mass to a bulk scalar zero mode that has been tuned to produce a massless 4-dimensional particle.  
 We use separation of variables to write 
\be
\psi(z) \phi(p_\mu) &\equiv&  \Phi(z,p_\mu)
\ee
This gives a Schrodinger equation for $\psi(z)$; the solutions are generally Bessel functions, but when the zero mode $\psi_0$ is massless in $4$-d we find a simple power-law in $z$.   Let us now ask how the $4$-d mass of this mode changes under a shift of the IR boundary condition.

Recall that the presence of a boundary mass term 
\be
S_{\rm bd} = -  \int d^4 x \sqrt{-g} 2  b k \Phi^2 
\ee
plus the canonical quadratic action in the bulk implies the following boundary condition on the wavefunction $\psi_0$ for $\Phi_0$, the lowest energy mode of $\Phi$:
\be
(\partial_y - b k ) \psi_0|_{\rm bd} &=& 0
\label{eq:scalarbdcond}
\ee
Following convention, we have defined the proper distance coordinate $y$ through $k z = e^{k y}$.  An IR (UV) localized massless zero mode has $b=\Delta$ ($b=4-\Delta$), respectively. The IR localized wavefunction is proportional to $(kz)^\Delta = e^{\Delta ky}$, and we can work out the normalization from
\be
\frac{1}{2} = \int_{y_{UV}}^{y_{IR}} dy (e^{-k y} \psi_0(y))^2
\ee
from which we learn that  in the limit where $y_{UV} \rightarrow -\infty$
\be
\psi_0(y_{IR}) = \sqrt{k} e^{k y_{IR}} \sqrt{\Delta-1}
\ee
Thus, if we take $b\rightarrow b+ \delta b$ in $S_{\rm bd}$ and evaluate the new contribution to $S_{\rm bd}$ from $\delta b$, we find
\be
\delta S_{\rm bd} = -2 (\delta b) \frac{1}{L_{IR}^2} (\Delta-1) \phi_0^2(p_\mu)
\ee
where we have used $\sqrt{g} =e^{-4 k y}$, $L_{IR}$ is the IR brane length scale $L_{IR} = e^{k y_{IR}} R_{AdS}$, and $\phi_0$ is the zero mode, so that the bulk field $\Phi \supset \psi_0(z) \phi_0(p_\mu)$.  Thus, we see that changing the IR boundary condition changes the $\phi_0$  mass by
$ \delta m^2 \sim  \frac{\Delta}{L_{IR}^2} $
for generic detunings of the boundary condition.  Massless scalar modes can only exist as a consequence of a symmetry or a tuning; moreover the tuning must be increasingly severe for larger $\Delta$.  

Now let us consider the general case and illustrate some features that arise at large dimension $\Delta$.  We will study how the $4$-d mass of the lightest KK mode in a large $\Delta$ bulk scalar field depends on the IR boundary conditions.  If we factor out $e^{\frac{3}{2}k y}$ from the bulk scalar wavefunction $\psi_0$ to define $g_0 \equiv e^{-\frac{3}{2} k y} \psi_0$, then the equation of motion can be written as
\be
g_0''(z) + \left( m_0^2 - \frac{4\Delta(\Delta-4)+15}{4 z^2} \right) g_0(z) &=&0,
\label{eq:geom}
\ee
which has only a single solution that is regular in the UV at $z=0$:
\be
g_0(z)&=& N \sqrt{z} J_{\Delta-2}(m_0 z).
\ee
The boundary condition (\ref{eq:scalarbdcond}) reads $L_{IR}g_0'(L_{IR}) +(\frac{3}{2}-b)g_0(L_{IR})=0$, which at large $\Delta, m_0 L_{IR}$ has the approximate numerical solution 
\be
m_0^2 \approx \frac{2}{L_{IR}^2 }   (\Delta-b)(\Delta-1) \sim \frac{2 \Delta^2}{L_{IR}^2}
\ee
 This agrees with our earlier calculation for $b \approx \Delta$, but it shows that in general $m_0^2 \propto \Delta^2 $ in the absence of tuning or symmetries.
 
  A general point that will be relevant for our later analysis is that, in the absence of any tuning of the boundary term $b$, we could have read off that $m_0^2 \sim \Delta^2$ immediately from the equation of motion (\ref{eq:geom}).  The reason is that (\ref{eq:geom}) has the form of a Schrodinger equation with a potential containing a $\sim \frac{\Delta^2}{z^2}$ term at large $\Delta$.  In a hard wall model, this pushes the zero mode up against the hard wall at $z=L_{IR}$, and the potential therefore gives a contribution to the ``energy'' (in this case, the eigenvalue $m_0^2$) 
 \be
 m_0^2 \sim \left\< \frac{\Delta^2}{z^2}\right\> \sim  \frac{ \Delta^2}{L_{IR}^2}. 
 \ee
Hard wall models lead to a conventional notion of decoupling, because high dimension operators in the broken CFT create heavy modes.
 
  In more general models of conformal symmetry breaking, we will continue to see this type of balance in the Schrodinger equation:   the bulk mass term $\Delta^2/z^2$ pushes modes towards larger $z$,  while the contributions from the dynamics of confinement pushes them toward smaller $z$.

\subsection{Chiral Fermions}

In this section, we discuss a case that involves the second mechanism of decoupling, through the small wavefunction overlap of heavy bulk modes with light degrees of freedom.

Specifically, in the case of fermions, zero modes can be protected by chirality (see e.g. \cite{Grossman:1999ra,ArkaniHamed:2001is,Hanson}).  In the presence of $n$ exact chiral symmetries, there will naturally be $n$ massless fermions in the $d$-dimensional broken CFT.  We will show that these massless fermions will be contained within the bulk fermion fields with small bulk masses whenever possible: if the fermion fields with small bulk masses outnumber the chiral symmetries, then all of the massless $d$-dimensional fermions will reside in the zero modes of these light bulk fields.    To see how this works in practice,  first recall the quadratic action for bulk fermions:
\be
S &=& \int  d^4 x \int_0^{z_{IR}} dz \sqrt{-g} \left( \bar{\Psi} i \partial^A \Gamma_A \Psi -M \bar{\Psi} \Psi  \right).
\label{eq:fermionaction}
\ee
Imposing a chiral symmetry on the fermions implies that the mass term $M$ should be interpreted as changing sign at the orbifold point $z_{IR}$ \cite{Grossman:1999ra,ArkaniHamed:2001is,Hanson}, and the isometries of pure AdS forbid a term of the form $\bar{\Psi} \Gamma^z \Psi$. 
 Bulk fermions are KK decomposed as
\be
\Psi_{L,R} &=& \sum_n \psi_n^{L,R}(x) z^2 k^2 f_n^{L,R}(z).
\ee
The KK wavefunctions $f_n$ satisfy \cite{Grossman:1999ra}
\be
\left( \pm  z \frac{\partial}{\partial z} - \frac{M}{k} \right) f_n^{L,R}(z) &=& - m_n z f_n^{R,L}(z) \nn\\
\int_0^{z_{\rm IR}} dz f_n^{L,R*} f_m^{L,R} &=& \delta_{mn} .
\label{eq:Psieom}
\ee
 A massless zero mode $(m_0=0)$ satisfies a simple first order equation. The zero mode for $L$ or $R$ is generally IR localized and therefore normalizable, and the boundary condition (reviewed below) that arises from the change in sign of $M$ at $z=z_{IR}$ is naturally consistent with the $m_0=0$ solution to the wave equation.  The lightest mode is consequently a massless zero mode.

What does this mean for decoupling of heavy bulk fields?  By imposing chirality for some fixed number of fermions, we can protect their masses from any corrections as we dial the bulk masses of the 5d fermions.  Clearly, in the limit that {\it all} the bulk fermions are taken to be very heavy, we will obtain a theory where decoupling fails, since some high dimension CFT operators will have a large overlap with massless 4d modes.  This occurs in this case because the fermion masses are protected by the chiral symmetry.

However, decoupling will happen if the number of chiral modes is no greater than the number of light bulk fields.  
We are then free to take all the remaining bulk fields in the theory to be heavy, and their corresponding 4d modes will become massive  along with them.  This is true even if we choose to lift the bulk masses of fermions that originally had large overlap with the massless 4d modes.  The reason is that conformal symmetry breaking generically mixes the different bulk fields, so that the massless 4d modes will generically obtain some overlap with all of the light bulk fermions.  These overlaps will become the dominant ones as the heavy bulk fermions are lifted to very large bulk masses.  

We will demonstrate this effect in a simple case, where mixing between two bulk fermions occurs only on a hard IR brane:
\be
S&=& \int  d^4 x \int_0^{z_{IR}} dz \sqrt{g} \left( \sum_{a=1}^2 \bar{\Psi}_a i \partial^A \Gamma_A \Psi_a - M_a \bar{\Psi}_a \Psi_a \right) + \int d^4 x \sum_{a,b=1}^2 \mu_{ab} \bar{\Psi}_a \Psi_b .
\ee
As above, the mass term $M_a$ should be interpreted as changing sign across $z=z_{IR}$ so that it preserves the chiral symmetry, whereas the explicit brane mass term $\mu_{ab}$ breaks this symmetry.  Due to the IR mass term, the bulk fermion wavefunctions can contain overlap with any of the KK modes:
\be
\Psi_a &=& \sum_n   \psi_{n}^{L,R}(x)z^2 k^2  f_{n,a}^{L,R}(z),
\ee
where the sum on $n$ is over all KK modes.  The equation of motion in the bulk is still
\be
 \left( \pm z \partial_z - M_a/k \right) f_{n,a}^{L,R}(z) &=& - m_{n} z f_{n,a}^{R,L}(z)
 \label{eq:fermionbulkeom}
\ee
but now subject to the boundary conditions \cite{Grossman:1999ra}
\be
0&=&\left. \left[ \left( \pm z \partial_z - M_a/k \right) \delta_{ac} + \mu_{ac}/k \right] f_{n,c}^L  \right|_{z=z_{IR}} , \nn\\
0&=& \left. f_{n,a}^R\right|_{z=z_{IR}},
\ee
and normalization conditions
\be
\sum_{a=1}^2 \int_0^{z_{\rm IR}} dz f_{n,a}^{L,R*} f_{m,a}^{L,R} =\delta_{nm}.
\ee
We are interested in the zero modes, $f_{0,a}$.
Let us take $\mu_{22}=\epsilon^2 \mu_{11}$ and $ \mu_{12}=\mu_{21}=\epsilon \mu_{11}$,  so that $\det(\mu_{ab})=0$, but we are taking $\mu_{11} \neq 0$.  Thus there is a massless zero mode for any $\epsilon$, and as the mixing parameter $\epsilon\rightarrow 0$ it sits completely
inside the heavy bulk field $\Psi_2$.   The massless zero mode has $m_n=0$ and thus its equation of motion (\ref{eq:fermionbulkeom}) can immediately be solved to be 
\be
f^L_{0,a}(z) &=& \frac{1}{{\cal N}^{1/2}} z^{M_a/k} y_a,  \qquad  {\cal N} = \left( \sum_a \frac{y_a^2 z_{\rm IR}^{2(M_a/k)+1}}{2(M_a/k)+1} \right).
\ee
where ${\cal N}$ is a normalization factor and the $y_a$ determine how the massless mode sits within the bulk modes.
The boundary condition just becomes $\sum_c \mu_{ac} y_c z_{\rm IR}^{M_c/k} =0$, so $y_c = \{-\epsilon z_{\rm IR}^{-M_1/k}, z_{\rm IR}^{-M_2/k} \}$. 
 Consequently, the overlap of the massless mode with $\Psi_2$ at large $M_2$ is
\be
|f_{0,2}^L(z)|^2 &\stackrel{M_2\gg k}{\longrightarrow}& \frac{k+M_1}{2\epsilon^2 M_2} \delta(z-z_{\rm IR}),
\ee
which vanishes at $M_2 \rightarrow \infty$. So we see that, for any non-zero $\epsilon$, as $M_2$ is taken  sufficiently large, the massless zero mode sits dominantly in the lighter bulk fermion, consistent with the new notion of decoupling.

\section{ Soft Wall Models and Effective Field Theory }
\label{sec:GeneralWall}

In this section we discuss the relationship between the $d+1$ and $d$-dimensional effective field theories in a more general context.  As in the case of the hard wall models discussed in section \ref{sec:HardWall}, we will find that high mass states in the warped bulk theory decouple from the dynamics of the light $d$-dimensional particles.   For the more general theories we consider here, this decoupling directly depends on bulk locality in a way that we explain in section \ref{sec:GeneralDecoupling}.

We study spacetimes with generalized warp factors that are asymptotically AdS in the UV region, but that break conformal invariance with a `soft wall' at a single scale of IR conformal symmetry breaking.  Thus we write the metric as
\be
\label{eq:WarpedMetric}
ds^2 = C(z)^2 (dz^2 + \eta_{\mu \nu} dx^\mu dx^\nu)
\ee
and assume that $C(z) \to \frac{1}{kz}$ as $z \to 0$.  We also assume that the metric obeys the Null Energy Condition (NEC); this is well-motivated in the bulk, and it has also been related to the $c$-theorem \cite{Zamolodchikov:1986gt} or $a$-theorem \cite{ Komargodski:2011vj, Komargodski:2011xv} in the CFT \cite{Freedman:1999gp,Girardello:1998pd,Myers:2012ed}.

We also allow for a general bulk dilaton to the theory, with a profile $e^{-\Phi(z)}$ multiplying the action for the other fields.  This can be motivated from several perspectives.  We expect that if conformal symmetry is explicitly broken by a running coupling, then there must be a bulk field with a spacetime dependent profile that breaks the AdS isometries down to the Poincar\'e group.  In fact, if there is a single running coupling $\alpha(\mu)$ then, via a field redefinition we can write the action of the broken CFT as
\be
S_{CFT} = \frac{1}{\alpha} \int d^d x \ \! \mathcal{L}(x)
\ee
This is probably the setup of greatest interest for us because we want to study theories that generate a single IR scale.  From the perspective of the holographic RG, where the $z$ coordinate roughly corresponds to the RG scale $1/\mu$, we expect that the overall normalization of the action must change with $z$ due to $\alpha(\mu)$, justifying an overall $e^{-\Phi(z)}$ coefficient in the $d+1$-dimensional action.  Furthermore, in many explicit top-down and bottom-up models of broken CFTs \cite{Klebanov:2000hb, Polchinski:2000uf, Kruczenski:2003be, Sakai:2004cn, Sakai:2005yt, Karch:2006pv, Katz:2007br, Batell:2008zm} there is either an explicit dilaton field or an analog that arises from the geometry of additional compact dimensions.  

This means that we will be studying an action of the form
\be
\label{eq:GeneralMatterAction}
S &=& \int d^d x  \int_0^\infty dz \ \! e^{-\Phi(z)} \sqrt{-g} \left( R - \Lambda + \mathcal{L}_{X}(g_{\mu \nu}, X_i) \right)
\ee
where the bulk fields $X_i$ are dual to CFT operators $\CO_i$.  $\mathcal{L}_{X}$ has the form of a standard EFT lagrangian with an expansion in powers of the bulk energy cutoff, corresponding to a cutoff on operator dimensions in the effective CFT.  We will not be solving Einstein's equations; instead we will examine general $C(z)$ and $\Phi(z)$ satisfying the NEC and leading to a broken CFT.

\subsection{ Bulk Wavefunctions in Soft Wall Models and the NEC}
\label{sec:NEC}

Let us begin by studying the behavior of free massive scalar fields in the background indicated in equation (\ref{eq:GeneralMatterAction}).  
The relevant part of the action is
\be
S &=& \int d^d x dz \ \! e^{-\Phi(z)} \sqrt{g} \left(  |D X |^2 - M^2 | X |^2 \right)
\ee
The equation of motion for $X$ can be solved by separation of variables between the $z$ and the transverse directions.  If we introduce the rescaled field $\psi$ and go to momentum space for the transverse directions, we can use the decomposition
\be
\label{eq:SeparationVariables}
\psi(z) \chi(p_\mu) &\equiv& e^{-B(z)/2} X(z,p_\mu), \ \ \ \mathrm{where} \ \ \ B(z) \equiv \Phi(z) -(d-1) \log C(z) .
\ee
The modes $\psi_n$ have the equation of motion
\be
-\psi_n'' + V(z) \psi_n &=& m_n^2 \psi_n \ \ \ \mathrm{with} \ \ \ p^2 = m_n^2,
\ee
and the Schrodinger potential $V(z)$ takes the form
\be
V(z) &=& \left( \frac{B'(z)}{2} \right)^2 - \frac{1}{2} B''(z)  + M^2 C^2(z) .
\label{eq:schr}
\ee
Solving the Schrodinger eigen-system gives the masses and bulk wavefunctions of the `mesons and glueballs' of the Broken CFT.  The dynamics will be largely controlled by the balance between different terms in equation (\ref{eq:schr}). In pure AdS, $C(z) \propto 1/kz$, so the bulk mass term makes a dominant contribution of $\frac{M^2}{(kz)^2}$ in the small $z$ region, where the spacetime is asymptotically AdS.  

\begin{figure}
\begin{center}
\includegraphics[width=0.7\textwidth]{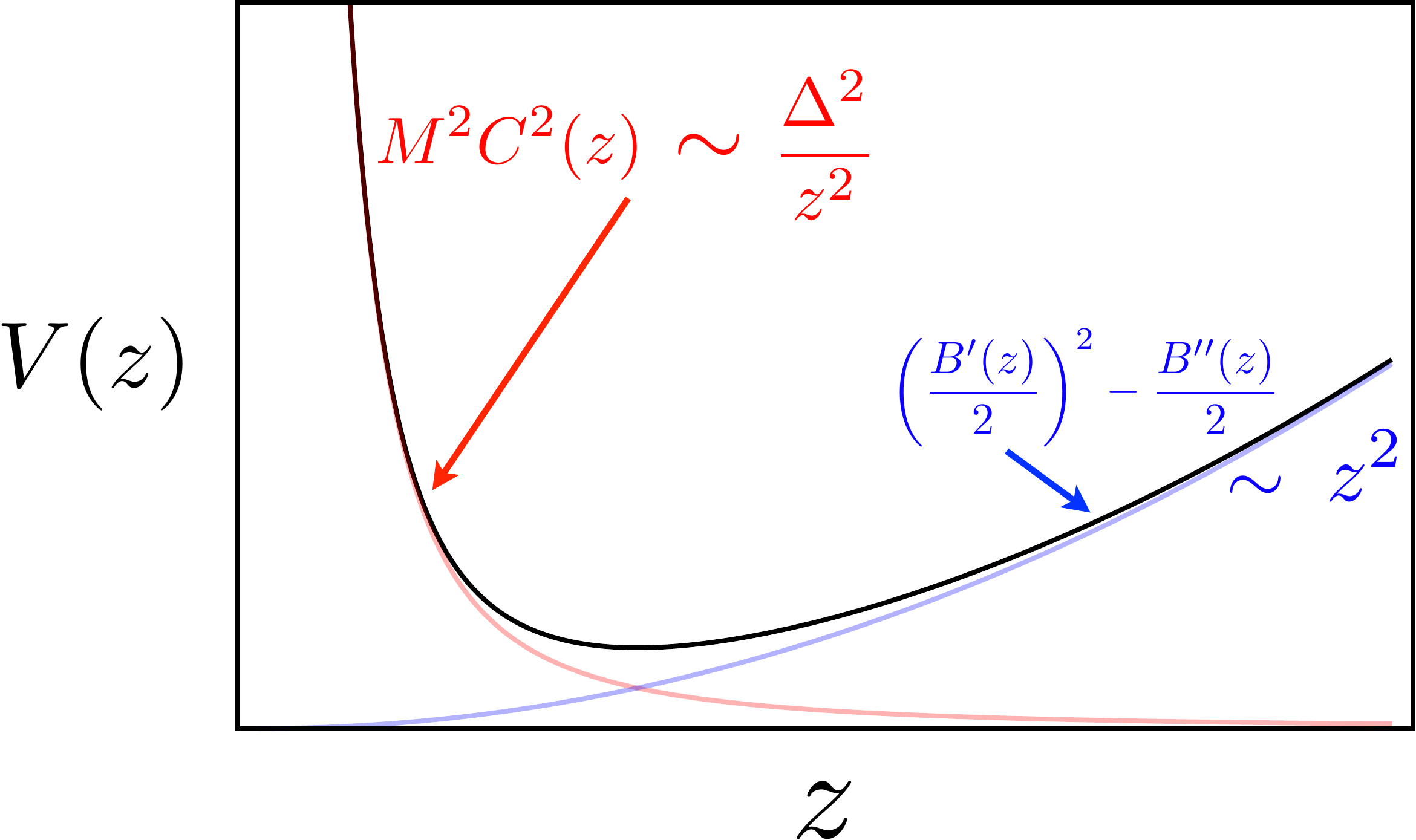}
\end{center}
\caption{The Schrodinger potential in equation (\ref{eq:schr}) for the KK mode wavefunctions gets contributions from the bulk mass as well as from the deviations from AdS due to the soft wall in the IR.  The former dominates at small $z$, behaving like $\frac{\Delta^2}{z^2}$ and pushing the wavefunctions toward larger $z$, whereas the latter tend to push the wavefunctions toward smaller $z$.  We have shown an example where the dilaton profile $\Phi(z)$ in equation (\ref{eq:SeparationVariables}) behaves like $\sim z^2$, resulting in a potential  $\propto z^2$ at large $z$, but the pattern is  general.  Modes are contained in a finite-sized `cavity' in the bulk, with a central $z$ that tends to grow with $\Delta$.
\label{fig:schro-potl}}
\end{figure}

We have seen that the behavior of bulk modes will be dictated by the potential $V(z)$, which depends on the function $C(z)$ in the metric and also on the dilaton profile.  Let us review constraints of the Null Energy Condition (NEC) on the warped metric in equation (\ref{eq:WarpedMetric}). Our formulae will be valid for AdS in any number of dimensions.  
We study general metrics consistent with two assumptions:
\begin{itemize}
\item The metric is approximately AdS in the `UV region', where $z \to 0$, so we have $C(z) \to \frac{1}{k z}$ in this limit.  
\item The energy-momentum tensor derived from the metric satisfies the Null Energy Condition (NEC) everywhere.  
\end{itemize}
Starting with the metric (\ref{eq:WarpedMetric}), one obtains the Einstein tensor
\be
G^\mu_{\ \nu} &=& (d-1) \delta^\mu_\nu \left( \frac{C''(z)}{C^3(z)} + \frac{d-4}{2} \frac{C^{\prime 2}(z)}{C^4(z)} \right), \nn\\
G^z_{\ z} &=& (d-1) \frac{d}{2} \frac{C^{\prime 2}(z)}{C^4(z)}, \nn\\
R &=&  \frac{-2}{d-1} \left( G^z_{\ z} + G^\mu_{\ \mu}\right).
\ee
The NEC implies that the quantity $G^z_{\ z} - G^0_{\ 0}$ is non-negative, so
\be
2\frac{C'(z)^2}{C^2(z)} \ge \frac{C''(z)}{C(z)}.
\ee
This implies that the function \cite{Myers:2010tj} 
\be
F(z) = - \frac{C^2(z)}{C'(z)}
\label{eq:FvsC}
\ee
must be non-increasing.  This function has been related to the central charge $c$ in $d=2$ \cite{Zamolodchikov:1986gt} and to $a$ in $d=4$ dimensions \cite{Komargodski:2011vj}, and in general dimensions is related to the holographic $c$-functions $a_d(z)$ defined in \cite{Freedman:1999gp,Girardello:1998pd,Myers:2012ed} by
\be
F^{d-1}(z) &= \ell_{\rm pl}^{d-1}  \left(\frac{\Gamma(d/2) a_d(z)}{\pi^{d/2}}\right).
\ee
In the next subsection, we consider constraints on soft wall models given the restriction above.  However, we note that our analysis does not apply to metrics of the form AdS$\times X$.  As far as we are aware, the existence and behavior of holographic c-functions has not been investigated for AdS $\times X$ compactifications where the metric of the compactification manifold $X$ changes with $z$.  In fact, we have observed that relevant metrics in the literature \cite{Karch:2002sh, Kruczenski:2003be} have NEC violation in the $(d+1)$-dimensional warped space-time geometry that results from a formal KK reduction on $X$.\footnote{Whether or not such a KK reduction gives rise to a sensible EFT is a separate question that requires further investigation.}  This suggests that after KK reduction, the current formulation of holographic c-theorems can break down.  It would be interesting to formulate a new holographic c-theorem for AdS $\times X$ in future work.

\subsubsection{Either a New IR CFT or a Wall That Ends Space}

We will now show that given the assumptions of the previous subsection, we must either recover a new asymptotically AdS metric as $z \to \infty$ or the space must end at large $z$ with the breakdown of the low-energy EFT.    Setups with growing curvature at large $z$ are rendered finite by the presence of new degrees of freedom or extra dimensions that resolve the singularity \cite{Klebanov:2000hb, Polchinski:2000uf, Karch:2002sh, Kruczenski:2003be, Sakai:2004cn, Sakai:2005yt}, resulting in a setup reminiscent of RS models with an `IR brane'.  

To understand why these conclusions follow, note that the $zz$ component of the Einstein tensor is
\be
G^z_{\ z} &=& -\frac{d(d-1)}{2F^2(z)}.
\ee 
Thus, in order to avoid large curvature regions with $|G^z_{\ z}|$ greater than some value $G_*$ where the bulk effective theory would break down, $F$ must never decrease below the critical value $F_*= \sqrt{\frac{d(d-1)}{2G_*}}$.  Moreover, since $F$ is non-increasing, this requires it to asymptote to its greatest lower bound $F_\infty$.  The only solution to equation (\ref{eq:FvsC}) with $F(z)$ approaching a positive constant $F_\infty>0$  as $z\rightarrow \infty$ is $C(z) \rightarrow \frac{F_{\infty}}{z}$, which is therefore asymptotically AdS in the IR as well as in the UV.

Curvature singularities on the other hand imply that the bulk space-time ends, in the sense that regions beyond the singularity depend on UV physics that can be encapsulated in some sort of boundary conditions at the singularity.  Extra compact dimensions may play a crucial role \cite{Klebanov:2000hb, Polchinski:2000uf, Sakai:2004cn, Sakai:2005yt}.  

However, models can still act like ``effective soft wall'' models, i.e. they look like soft wall models for many of the light KK modes, with only heavy KK modes are affected by the hard wall boundary conditions.  In particular, if the Schrodinger potential (\ref{eq:schr}) experienced by the bulk fields rises sharply enough as the singularity is approached, then the low-lying KK modes would have to tunnel through a potential barrier to feel the hard wall.  To a good approximation these modes would behave like the modes of a soft wall model.  

To see this intuition borne out in detail, we can write the contribution to the potential $V$ in equation (\ref{eq:schr}) from the metric\footnote{In this subsection, we are considering the case where the dilaton profile is absent, so $\Phi=0$ in equation (\ref{eq:schr}), and we are also using the general $d$ formula for $B$, so $B(z) \equiv  (1-d)\log C(z) $.} as 
\begin{equation}
V_C(z)\equiv (d-1)\left(  \frac{d-3}{4}  \frac{(C')^2}{C^2} + \frac{1}{2} \frac{C''}{C} \right) =
-\frac{(d-1)C^2}{4d} R =
 \frac{C^2}{2d} \left( G^z_z +G^\mu_\mu \right),
\end{equation}
where $R$ is the geometric scalar curvature.
If there is no cancellation in the potential $V_C$ between $G^z_z$ and $G^\mu_{ \ \mu}$, we can analyze its divergence behavior by considering just the $G^z_{\ z}$ component.\footnote{In fact, keeping track of $G^\mu_{\ \mu}$ would not change our conclusions, though it makes the argument slightly more complicated.  For $C(z) \nrightarrow 0$, curvature singularities in $G^\mu_{\ \mu}$ must be accompanied by curvature singularities in $G^z_{\ z}$, because $G^\mu_{\ \mu}$ is a linear combination of $\frac{1}{F^2(z)}$ and $\frac{1}{C(z)} \partial_z \frac{1}{F(z)}$; since $\frac{1}{F^2(z)}$ is monotonically increasing, its derivative cannot diverge without its value diverging. But our argument shows that singularities in $\frac{1}{F^2(z)}$ do not occur if $C(z) \nrightarrow 0$.  Therefore singularities in either $G^z_{\ z}$ or $G^\mu_{\ \mu}$ require $C(z) \rightarrow 0$.} 
Neglecting numeric constants, we therefore have
\be
V \sim \left( \frac{C'}{C}\right)^2 , \qquad G^z_z \sim \left( \frac{C'}{C^2}\right)^2.
\ee
Consequently, singularities in the curvature occur at 
\be
\frac{C'}{C^2} \rightarrow \infty,
\ee
whereas singularities in the potential occur at
\be
\frac{C'}{C} \rightarrow \infty .
\ee
Both of these require $C\rightarrow 0$ (positivity of $F(z)$ implies that $C'$ is negative, so it cannot grow to $\infty$ without passing through a singularity at a smaller value of $z$).  Since $V_C \propto C^2 R$, at $C\rightarrow 0$, it is possible to have curvature singularities without potential singularities, but it is not possible to have potential singularities without curvature singularities.  Essentially, what is necessary to achieve such ``effective soft-wall'' setups is that the curvature asymptotically in the UV must be small (compared to the Planck scale), so that $G^z_{\ z}$ has a long way to grow before reaching large curvature.  This allows the Schrodinger potential to become very large at the singularity and thus push the light KK modes  away.  It is likely not possible to have true soft-wall models, where all KK modes would have to tunnel through an infinite potential in order to feel the singularity.

\subsubsection{Examples}
\label{sec:Wavefunctions}

Let us consider two concrete examples that exhibit ``hard-wall'' and ``effective soft-wall'' behavior.  We will take $d=4$ to avoid clutter. The first is
\be
C(z) = \frac{1}{kz(1+zT)}
\ee
which has $V_C\rightarrow 0$ at $z\rightarrow \infty$, but the geometric scalar curvature $R(z) \rightarrow -(8k z T)^2$ at large $z$.  Thus,  both terms in the potential push the modes in the same direction and consequently the potential does not push modes away from the singularity at all. Thus the theory is a hard-wall model.\footnote{Including the mass term $M^2 C^2(z)$ in equation (\ref{eq:schr}), the full potential is $V(z) = \frac{4 (M/k)^2 +15+48 zT (1+zT)}{4 z^2(1+zT)^2}$, so for $(M/k)^2>-15/4$, the potential is strictly decreasing everywhere and all modes feel the hard wall.}

The second example is  
\be
C(z) &=& \frac{1-z T}{kz}.
\ee
This gives 
\be
V_C(z) &=& \frac{3(5-4 z T)}{4z^2 (1-z T)^2}, \qquad \qquad R(z) = \frac{ 4k^2(4zT-5)}{ (1-zT)^4} .
\ee  
The potential therefore blows up at $z\rightarrow T^{-1}$ at a rate that is only power-law slower than the rate at which the curvature blows up. Consequently, the value of $k$ compared to the cut-off of the theory determines how  many light KK modes are pushed away from the curvature singularity.

Now let us consider a canonical example for the bulk wavefunctions, where we can solve the Schrodinger equation (\ref{eq:schr}) explicitly.  The model we consider  \cite{Katz:2007br}, which assumes an everywhere AdS metric and dilaton profile $\Phi(z) = a z^2$ yields a linear spectrum.  For simplicity we now stick to $d=4$; we find a Schrodinger equation
\be
\label{eq:LinearSchrodinger}
-\psi_n'' + \left[ \frac{4 M^2 + 15}{4 (kz)^2} + a^2 z^2 \right] \psi_n &=& (  m_n^2 - 2a) \psi_n 
\ee
This has eigenvalues for the masses
\be
\label{eq:LinearConfinementMasses}
m_n^2 = a \left( 2m + 2 + 4n \right)  \ \ \ \mathrm{where} \ \ \ m \equiv \sqrt{(M/k)^2 + 4} 
\ee
 The wave functions are 
\be
\label{eq:LaguerreWavefunction}
\psi_{n}(z) = \sqrt{\frac{2 n!}{(m+n)!}} e^{-z^2/2} z^{\frac{1}{2} + m}  L_m^n(z^2)
\ee
where we recall that $m \equiv \sqrt{(M/k)^2 + 4} $ in terms of the bulk mass $M^2 = k^2 \Delta(\Delta-d)$, where $\Delta$ is the dimension of the dual CFT operator.  The functions $L_a^n(z^2)$ are Laguerre polynomials of degree $2n$ in $z$.  
We see that at both large $M \approx k \Delta$ (where $\Delta$ is the dimension) and at large $n$, the squared masses are linear in $\Delta$ and $n$, as expected for QCD.   With this choice of metric and dilaton we also have a linear spectrum of squared masses $m_\ell^2 \propto \ell$ for higher spin bulk fields at large spin $\ell$.  This result depends in detail on the choice of metric and dilaton profile, because higher spin fields experience the metric and dilaton in a different combination because of their tensor indices.  In fact it was shown in \cite{Katz:2007br, Karch:2010eg} that this choice for the dilaton is the unique choice that gives a linear spectrum.

In the example above, we saw that the approximately AdS region of the metric in the UV causes $V(z)$ to grow as $z \to 0$, while the soft-wall potential from the dilaton causes $V(z)$ to grow at large $z$, pushing the bulk wavefunctions away from this region.  Since the contribution to the potential at small $z$ grew as $M_{bulk}^2$, in the case of larger bulk masses (CFT operator dimensions) the wavefunctions were localized at larger values of $z$.   This pattern also holds in the general case, as we see in figure \ref{fig:schro-potl}.  The asymptotically AdS metric always causes the potential to grow as $z \to 0$, pushing the meson wavefunctions towards large $z$, with a larger force for larger values of the operator dimension $\Delta$.  The potential from the dilaton and/or the hard-wall pushes meson wavefunctions towards smaller values of $z$, containing them in some finite sized cavity in the $z$ direction.

\subsubsection{Soft Wall Fermions}

The situation with chiral fermions in soft wall models is only slightly modified from the hard wall case.  There is no boundary brane on which chiral symmetry may be broken, and the usual AdS-invariant quadratic action (\ref{eq:fermionaction}) produces an equation of motion (\ref{eq:Psieom}) that decouples left- and right-handed modes for zero KK mode mass $m_0$, so naively there is always a massless zero mode.  However, the breaking of chirality and scale symmetry allows additional bulk mass terms of the form
\be
S = \int d^5 x \sqrt{g}  \mu_5(z) \bar{\Psi} \Gamma^z \Psi.
\ee
Since $\Gamma^z$ is proportional to the usual $\gamma^5$ matrix, non-vanishing $\mu_5(z)$ mixes left- and right-handed 4d fermion modes, and therefore produces a massive zero mode.  Chiral symmetry could still forbid such a term, naturally leading to a massless zero mode solution to the fermion wave equation.

\subsection{Decoupling of High Dimension Operators from the 4-d EFT}
\label{sec:GeneralDecoupling}

 In the first decoupling mechanism we stated in the introduction,  decoupling is a formal way of saying that we can describe the light particles with a low-energy effective field theory. The corrections from high energy states can be encoded in irrelevant operators that make small contribution to low-energy observables.  

The second type of decoupling arises naturally from AdS descriptions of CFTs with an IR breaking of conformal invariance.%, is the decoupling of operators with a large scaling dimension from the interactions of the low-dimension spectrum.
  This is the natural form that decoupling takes if we consider the dilatation operator $D$ as the Hamiltonian of the system.

We will now give a very rough but rather general argument explaining why in broken CFTs, the states created by large dimension operators naturally decouple from the low energy states.  Our argument in this section will be based on a Poincar\'e invariant warped compactification, so formally it requires some sort of large $N$ expansion for the broken CFT.  

There are two possible circumstances that can arise if we have only a single IR scale:
\begin{enumerate}
\item All modes are localized in the same region of the bulk, because all terms in the effective Schrodinger potential are proportional to $\Delta^2$.  In this case,  the position of the potential at the minimum will be independent of $\Delta$, and $m_{4-d}^2 \propto \Delta^2$ and decoupling is guaranteed, as in RS-type `hard wall' models.  
\item Otherwise, the effective Schrodinger potential can be written as $\frac{\Delta^2}{z^2} + V(z)$.  Now we still expect that $m_{4-d}$ will grow with $\Delta$, but these quantities need not be proportional.  For example, an extreme case would be  $V(z) \sim \log(1+z)$, so we would roughly have $m_{4-d}^2 \sim \log \Delta$; obviously the linearly confining soft wall is a more reasonable example.

In these cases the modes will not reside in the same region of $z$, and in fact the wave function overlaps between large $\Delta$ and small $\Delta$ will be exponentially suppressed, guaranteeing an even more powerful form of decoupling.  Let us now see this in detail.
\end{enumerate}

One can evaluate our last claim in many specific solvable models, but to make the argument in general we will make some approximations.  Let us focus only on zero modes in the bulk.  Highly excited modes will range over a large region in the bulk and will eventually mix, although their interactions will be diluted by volume factors.  

We will assume that in natural units the potential $V(z)$ has $O(1)$ coefficients, so that in particular mesons with $\Delta \sim 1$ will be localized near some $z_* \sim 1$.  The meson wavefunctions obey a Schrodinger equation that we can write as
\be
-\psi_{\Delta,n}'' + \left[ \frac{\Delta^2}{z^2} + V(z) \right] \psi_{\Delta, n} &=& m_{\Delta, n}^2 \psi_{\Delta, n} .
\ee
We have replaced $M^2/k^2 = \Delta^2 - 4 \Delta$ with simply $\Delta^2$ in order to avoid clutter; since we are mainly interested in the behavior at large $\Delta$, this will have no qualitative bearing on the following results. We will define $V_{tot}(z) \equiv \frac{\Delta^2}{z^2} + V(z)$ as the full potential.
The key physical point is that the $\Delta$ dependence comes from the AdS part of the metric, while $V(z)$ does not depend signficantly on $\Delta$.  This is what causes mesons with very different values of $\Delta$ to be localized in different regions in the bulk.  Note that if instead $V(z)$ was proportional to some increasing function of $\Delta$, then we would obtain meson masses $m_{\Delta,0}^2$ that manifestly grow with $\Delta$, leading to a traditional form of decoupling, where high dimension operators create high mass states.  

The zero modes of all mesons will be naturally centered at $z$ such that their effective potentials are minimized.  Thus for operators with large $\Delta$, the zero mode wave functions will be centered at $z_\Delta$ satisfying\footnote{At large $\Delta$ we expect that the full potential $V_{tot}$ will only have one minimum.}
\be
z_{\Delta} = \left( \frac{2 \Delta^2}{V'(z_\Delta)} \right)^{1/3}
\ee
We will approximate the zero mode wave functions for $\CO_\Delta$ with harmonic oscillator wave functions centered at $z_\Delta$ and $z_*$.  We justify this approximation in a fairly general class of models in appendix \ref{sec:GeneralDecoupling}.  We could perform a more detailed analysis in any specific case by either solving the system numerically or by using the WKB approximation.  Our approximation will give an approximate overlap function
\be
|\psi_*(z) \psi_{\Delta,0}(z)| \approx \exp \left[  -\frac{V_{tot}''(z_\Delta)}{4} (z-z_\Delta)^2 - \frac{V_{tot}''(z_*)}{4} (z-z_*)^2 \right]
\ee
where $\psi_*(z)$ is the bulk wavefunction for a meson created by an operator with order one dimension.  All of the $z_*$ dependent terms are $\CO(1)$ by assumption, so roughly speaking we can write the overlap as
\be
|\psi_*(z) \psi_{\Delta,0}(z)|  \approx \exp \left[  -\frac{V_{tot}''(z_\Delta)}{4} (z-z_\Delta)^2 - \frac{1}{4} z^2 \right]
\ee
  This overlap function will be maximized when the exponent is at a minimum, with
\be
z \approx \frac{z_\Delta V_{tot}''(z_\Delta)}{1 + V_{tot}''(z_\Delta)}
\ee
In this region we find an overlap
\be
|\psi_*(z) \psi_{\Delta,0}(z)|  \approx \exp \left[  -\frac{1}{4} z_\Delta^2  \left( \frac{ V_{tot}''(z_\Delta)}{1 + V_{tot}''(z_\Delta) } \right) \right]
\ee
The term in parentheses involving second derivates will be  $O(1)$ if $V(z)$ grows quickly,  but for  shallow $V(z)$ it can be small.  In general we have $V_{tot}''(z) \approx \Delta^2 / z^4$, so we find
\be
\label{eq:OverlapApprox}
|\psi_*(z) \psi_{\Delta,0}(z)|   \approx \exp \left[  -\frac{1}{4} z_{\Delta}^2 \left( \frac{\Delta^2}{\Delta^2 + z_\Delta^4} \right) \right]
\ee
This is the suppression factor $f(\Delta)$ from equation (\ref{eq:TwoPointFormFactor}) in the introduction.
 If we have an effective soft-wall, then we obtain an exponent that grows with $\Delta$ and therefore a very suppressed overlap between wave functions.  If we have a very hard wall, e.g.  $V(z) = e^z$, then $V'(z_\Delta) \sim \Delta^2$ will also be large, and there will not be much suppression of the wave function overlaps.  However, in that case we will have $m_{4-d}^2 \approx \Delta^2$, so that the contributions of high dimension operators will decouple because they create high mass states.  To emphasize that large $m_{4-d}^2$ is a fairly generic consequence of avoiding the exponential (\ref{eq:OverlapApprox}), recall that the mass-squared can be approximated by inspection of the Schrodinger equation:
\be
m_{4-d}^2 \sim \left\< \frac{\Delta^2}{z^2} \right\> + \< V(z) \> \sim \frac{\Delta^2}{z_\Delta^2} + V(z_\Delta).
\ee
Both terms on the RHS grow parametrically when $z_\Delta \sim 1$, the first manifestly so and the second because $z_\Delta \sim 1$ requires $V'(z_\Delta) \sim \Delta^2$.  A caveat is that, as we noted in section \ref{sec:HardWall}, boundary conditions may be finely tuned (or protected by symmetry) so that the contribution to the mass-squared from the potential is exactly canceled by the contribution from the kinetic energy term $-\psi_{\Delta,0}''$.

In this section we have used a rather coarse approximation, treating the meson wavefunctions as though they are harmonic oscillator wavefunctions, so the reader might wonder if specific models behave differently.  In the case of the linearly confining soft-wall model \cite{Karch:2006pv},  one can use the explicit form for the wavefunctions to show that $|\psi_{\Delta,0} (z) \psi_{*}(z)| \lesssim \Delta^{\frac{3}{2}} 2^{-\Delta /2}$ at large $\Delta$.  In these models $z_\Delta \approx \sqrt{\Delta}$, so our approximation in equation (\ref{eq:OverlapApprox}) has correctly predicted the power of $\Delta$ occurring in the exponentially suppressed wavefunction overlap.  We should also emphasize that this provides a concrete prediction for two-point functions in linearly confining theories such as QCD -- we would expect that $p=1$ in equation (\ref{eq:TwoPointFormFactor}).

In conclusion, we expect that high dimension operators will decouple, either via an ordinary power-law mass suppression, or due to a more interesting exponential form factor.  Furthermore, we expect faster exponential shutoff when the meson states $m_\Delta^2(n)$ are more closely spaced as a function of $n$.

\subsection{A Comment on Meson Sizes}
\label{sec:Sizes}

A familiar intuition from AdS/CFT posits that bulk objects localized at different values of the $z$ coordinate can be viewed as CFT states with a characteristic size set by $z$.  Taken literally, this would suggest that mesons localized at different values of $z$ in the bulk duals of broken CFTs might have very different characteristic sizes.  This intuition seems plausible if we take QCD as an example, where we would expect that heavy excited mesons correspond, very roughly speaking, to long strings or bound states of quarks with large orbits.  One might also have expected the size and $d$-dimensional wave functions of mesons to help explain our decoupling results from the previous section.  However, this reasoning does not seem to be borne out by a more careful analysis of the bulk theory.

Our treatment of holographic meson sizes will be simple and brief; for a more complete analysis that discusses many different measures of size see \cite{Hong:2003jm, Hong:2004gz}.  The idea is that we want to measure the transverse size of the meson by probing it with some other field, such as a scalar, a current, or the gravitational field.  In the case of a scalar probe $\CO_S$ on a scalar meson, we can write the matrix element in terms of a form factor
\be
\langle p +q | \CO_S(0) | p \rangle = F_S(q^2)
\ee
This form factor can be computed as usual in AdS/CFT if the bulk field $S$ dual to $\CO_S$ has a bulk 3-pt coupling with the meson, which for concreteness one can take to be
\be
\int d^4 x dz \sqrt{g} \lambda S X^2 ,
\ee
where $X$ is the bulk field dual to the mesons and $S$ is the bulk field dual to the operator $\CO_S$ being used to probe the meson size.  The value of the coupling $\lambda$ is irrelevant to the size of the mesons, so long as it is small enough that back reaction can be neglected.  
 One simply convolves the bulk-boundary propagator for $S$ with the meson wavefunction in the bulk.
 \be
F(q^2) =  \lambda \int_0^\infty dz \ \!  
G_{\partial B}^S(z, q) \left( C^2(z) \psi_{\Delta, n}^{in} (p, z) \psi_{\Delta, n}^{out} (p+q, z) \right)
\label{eq:formfactorintegral}
\ee
where $G_{\partial B}^S$ is the bulk-to-boundary propagator for $S$, and the $\psi_{\Delta, n}$ are the bulk wave functions for the $X$ mesons.
As discussed in \cite{Hong:2003jm}, in terms of the form factor $F(q^2)$ we can define a `transverse scalar charge distribution' as
\be
\tilde F_S(x_\perp^2) = \int d^{2} q_\perp e^{i q_\perp \cdot x_\perp} F_S(q_\perp^2)
\ee
where $p \cdot q_\perp  = 0$,  so the space-like momentum transfer $q_\perp$ is orthogonal to the meson momentum.  The characteristic scale on which $\tilde F_S(x_\perp^2)$ has support gives an estimate of the meson size, and in particular we can measure moments such as $\langle x_\perp^2 \rangle$.  This natural measure of size can be expressed as
\be
\label{eq:SizeFormula}
\langle r^2 \rangle = \left.  \frac{\frac{\partial}{\partial q^2} F_S(q^2)}{F_S(q^2)} \right|_{q \to 0}
\ee
Using this general method with a variety of probes, it was shown in \cite{Hong:2003jm} that in RS-type hard-wall models, all hadrons have roughly the same size.  This means that in broken CFTs dual to RS-type models, hadrons do not grow appreciably at large angular momentum or excitation number, nor is their size affected by the dimension of the CFT operator that creates them.

Let us consider soft-wall models.  For simplicity we will study the soft-wall model of \cite{Karch:2006pv} which produces a linear squared mass spectrum in excitation number and in meson angular momentum.  The bulk wave functions for scalar mesons obey the Schrodinger equation (\ref{eq:LinearSchrodinger}).  As a simple consequence of the Virial Theorem, the expectation value of the bulk coordinate $\langle z^2 \rangle$ grows linearly with both dimension $\Delta$ and with excitation number $n$ when these parameters are large.  However, the physical meson size $\langle x_\perp^2 \rangle $ does not grow in this way.  One can compute this in a parallel manner to the hard wall case.  The 3-pt bulk interaction now contains a factor of the dilaton profile:
\be
\int d^{d} x dz \ \! e^{- \Phi(z)} \sqrt{g} \left( \lambda S X^2 \right)
\ee
where as before we are using the bulk scalar field $S$ dual to a CFT operator $\CO_S$ in order to probe the mesons represented by the bulk $X$ modes.   
We need to fix the properties of $\CO_S$, so we will assume it has a fixed dimension $\Delta_S=d=4$, dual to a massless bulk field.\footnote{A larger dimension for $\CO_S$ would change our results for low-lying mesons, but it will not affect the limit of large bulk mass of $X$ or large excitation number of the mesons.}  In the soft wall model of \cite{Karch:2006pv} that produces a linear spectrum of meson masses the form factor is still of the form (\ref{eq:formfactorintegral}), 
but with $G_{\partial B}^S$ and $\psi_{\Delta,n}$ now computed in the presence of the dilaton profile background.   The resulting $X$ meson bulk wavefunctions $\psi_{\Delta, n}$ were already  given in equation (\ref{eq:LaguerreWavefunction}) but for simplicity we repeat that they are
\be
\psi_{\Delta,n}(z) = \sqrt{\frac{2 n!}{(m+n)!}} e^{-z^2/2} z^{\frac{1}{2} + m}  L_m^n(z^2)
\ee
where we recall that $m \equiv \sqrt{(M/k)^2 + 4} $ in terms of the bulk mass $M^2 = k^2 \Delta(\Delta-d)$, where $\Delta$ is the dimension of the CFT operator $\CO_X$.  
The functions $L_a^n(z^2)$ are Laguerre polynomials of degree $2n$ in $z$.  
These wavefunctions are normalized to include the metric and dilaton as in equation (\ref{eq:SeparationVariables}), so that 
\be
\int_0^\infty dz \ \!  |\psi_{\Delta, n}(z) |^2 = 1
\ee
To compute the parametric size of the mesons we need to know  the bulk-boundary propagator in the presence of the dilaton profile background
\be
G_{\partial B}^S(z, q)  \propto  z^{ m+2} \ \! U\left(\frac{q^2 L_{IR}^2 }{4}+\frac{m}{2}+1, m +1,z^2\right),
\ee
where $U$ is the confluent hypergeometric function, and $L_{IR}$ is the characteristic IR length scale.  

To compute the size of mesons we can use equation (\ref{eq:SizeFormula}).  Note that the size can only be extracted once we normalize the form factor so that the long-range force on the meson center of mass is fixed.  The effect of this normalization can be understood heuristically by studying the function
\be
\left. \frac{ \frac{\partial}{\partial q^2} G_{\partial B} (z, q) }{G_{\partial B} (z, q)} \right|_{q \to 0} \approx \frac{L_{IR}^2}{2} \log(z)
\ee
where the approximation holds at large $z$.  Because this function is slowly varying at large $z$, its expectation value  weighted by the squared meson wavefunction gives a good estimate of the meson size from equation (\ref{eq:SizeFormula}). This gives 
\be
\langle r^2 \rangle &\approx& \int_0^\infty dz \ \! 
\frac{L_{IR}^2}{2} \log(z) \left| \psi_{\Delta, n} ( z) \right|^2 
\\
& \approx & \frac{L_{IR}^2}{2}  \log \langle z_{\Delta, n} \rangle
\ee
This is the result -- the mesons only grow logarithmically with $\langle z_{\Delta, n} \rangle$.  We have $\langle z_{\Delta, n} \rangle \propto \sqrt{\Delta}$ at large dimension and $\langle z_{\Delta, n} \rangle \propto \sqrt{n}$ at large excitation number, so the meson size grows only asymptotically as $\log \Delta$ or $\log n$.  This result will not change if we consider more complicated bulk interactions, and we found it to be independent of the bulk mass $M_S$ of the probe field.  This universality follows from the normalization condition on the form factor in equation (\ref{eq:SizeFormula}).  The result also follows from dimensional analysis once we note that $q$ appears in the combination $q L_{IR}$ in the propagator $G_{\partial B}$.

We conclude that despite being well-separated in the bulk, mesons created by operators of very different dimensions do not have very different physical sizes in soft-wall models.  The proximate reason is that  highly excited states are localized in a region where the soft wall is badly breaking the AdS symmetries.

\section{Discussion}
\label{sec:Discussion}

CFTs can have two distinct Effective Field Theory descriptions.  The first, which we have called the $d$-dimensional EFT, is the standard Wilsonian one where momentum shells are integrated out along the RG flow.  The second notion of EFT for describing Conformal Field Theories is motivated by AdS/CFT duality. In this ``$(d+1)$-dimensional EFT'', one integrates out states based on their scaling dimension or conformal Casimir.  Since scaling dimensions are dual to bulk masses, in the $(d+1)$-dimensional EFT we simply integrate out heavy bulk fields when their mass is above the cut-off.  

A priori, these two very different EFTs are only loosely related.  In particular, when conformal symmetry is preserved, the spectral decomposition of an arbitrarily massive bulk field shows that it contains a continuum of light states down to zero mass.  After conformal symmetry is broken, however, $d$-dimensional modes obtain a discrete spectrum of masses, and CFT operators of different dimensions mix with each other.  It then makes sense to ask if and how quickly large dimension operators decouple from the lowest-mass states in the theory.  In this paper, we have studied conformal symmetry breaking and its consequences for the modes dual to heavy bulk fields in order to analyze the relationship between the bulk vs. boundary EFT constructions.  We find that there is a single stronger notion of decoupling where both the high dimension operators and the heavy states decouple from the low energy physics.

Our analysis makes use of the assumption of a local bulk dual, and along the way we have explored various possibilities for the realization of confinement through the warp factor of the bulk metric.  Constraints from the NEC have a well-known relation to holographic c-functions \cite{Myers:2012ed}, and their monotonicity limits the possible behavior of soft-wall models. We have shown that although no true soft wall models exist, effective soft wall models can constrain states away from infinity. 
It is notable that higher-dimensional metrics of the form AdS$_5\times X$ with additional compact dimensions in $X$ can satisfy the NEC and yet upon a KK reduction produce AdS$_5$ metrics that violate it when the size of $X$ varies with the AdS radial direction. % It would be interesting to explore this possibility further and find a generalization of the holographic c-function for AdS$ \times X$ spacetimes. 
%We have also noted the null energy condition is violated in the lower-dimensional theory when compactified dimensions are present.
 It will be interesting to see under which conditions this arises and whether a valid effective theory applies to these examples, as well as to find a generalization of the holographic c-function for AdS$ \times X$ spacetimes.

An ambitious goal of holographic studies has been to systematically develop $(d+1)$-dimensional holographic duals for strongly-coupled field theories whose UV conformal symmetry is broken in the IR.  Informally, we are trying to transform AdS/QCD and AdS/CMT from art to science.  Such constructions attempt to reproduce the low-lying spectrum of massive states in the broken CFT using the lightest fields in the bulk theory.  To justify this approach, one must assume that the light resonances have their dominant overlap with CFT operators of low dimension, or more precisely, of small conformal Casimir. Our analysis of the relation between the bulk and boundary EFTs  provides evidence for this assumption.  Moreover, we have seen that
the decoupling can often be exponential.   We find this to be an encouraging sign that an approximation which includes only the lowest dimension operators may be effective even in theories with a mild or non-existent gap
in the dimension of operators.  In particular, we note that our prediction for QCD is that the decoupling is exponential, a fact which is
confirmed by studies of 2D QCD at large N.

We have used the bulk space-time as an analytical tool, but it would be very useful to have an argument for the decoupling of high-dimension operators directly in the broken CFT.  In this way we could determine which assumptions concerning the broken CFT are necessary to imply the decoupling that we observe in holographic models.  Some examples include the assumptions of large $N$ and a small number of low-dimension single-trace operators \cite{JP, Heemskerk:2010ty,  ElShowk:2011ag, Sundrum:2011ic, AdSfromCFT}, and alternatively our assumption that there is only a single running coupling and $z$-dependent bulk profile.  Since the suppression of high-dimension operators in broken CFTs can already be seen by studying two-point functions, it might be interesting to study these questions on the lattice.

We have noted that the growth of the physical size of mesons and glueballs tends to be much slower than the growth in their average position $\< z\>$  in the bulk in the limit of large operator dimension or meson excitation number.  Consequently, although soft-wall bulk EFTs can reproduce the Regge spectrum of QCD, it seems that they do not reproduce expectations for the sizes of excited meson and glueball states \cite{Csaki:2008dt} based on the confining string picture.  It would be interesting to understand whether this can be resolved, and if a bulk string theory is necessary.

\section*{Acknowledgments}

We thank Simeon Hellerman, Shamit Kachru, Rob Myers, Joe Polchinski, and Matt Schwartz for discussions.
ALF and JK thank the GGI in Florence for hospitality while this work was supposed to be completed.  This material is based upon work supported in part by the National Science Foundation Grant No. 1066293. ALF was partially supported by ERC grant BSMOXFORD no. 228169. JK acknowledges support from the US DOE under contract no. DE-AC02-76SF00515.  The work of LR was supported in part by NSF grant PHY-0855591, NSF grant PHY-0556111 and the Fundamental Laws Initiative of the Harvard Center for the Fundamental Laws of Nature.   EK is supported by DOE grant DE-FG02-01ER-40676 and NSF CAREER grant PHY-0645456.

\appendix

\section{Justifying the Quadratic Approximation and \\ Relating $f(\Delta)$ to the Density of States}
\label{sec:JustifyingQuadratic}

In section \ref{sec:GeneralDecoupling} we provided some very general arguments that the $d$-dimensional EFT description follows from the $(d+1)$-dimensional warped EFT description.  Here we provide a slightly more general example and show why the arguments of section \ref{sec:GeneralDecoupling} apply to it.  Specifically, we will consider the Schrodinger equation
\be
\label{eq:ZAlphaPotential}
-\psi_{\Delta,n}'' + \left[ \frac{\Delta^2}{2 z^2} + z^{\alpha} \right] \psi_{\Delta, n} &=& m_{\Delta, n}^2 \psi_{\Delta, n} 
\ee
and show that for large $\Delta$ and any $\alpha > 0$, the wavefunctions are well-approximated by harmonic oscillator wavefunctions.  The point is very simple.  This potential has a minimum at
\be
z_\Delta = \left( \frac{\Delta^2}{\alpha} \right)^{\frac{1}{2 + \alpha}}
\ee
If we expand the potential about this point at quadratic order, we find a width
\be
\delta z = \frac{1}{\sqrt{V_{tot}''(z_\Delta) }} = \frac{ \Delta
   ^{\frac{4}{\alpha +2}-1}}{\alpha ^{\frac{2}{\alpha +2}} \sqrt{\alpha+2}}
\ee
Now if we consider the variation in the potential from the cubic and higher terms at $z = z_\Delta \pm \delta z$, we only find a change
\be
V_{tot}'''(z_\Delta) \delta z^3 = \frac{(\alpha -5) }{\alpha ^{\frac{1}{\alpha
   +2}}  \sqrt{\alpha +2} } \times \frac{1}{\Delta ^{\frac{\alpha }{\alpha
   +2}}}
\ee
Thus we see that for all $\alpha > 0$, as we study the limit of large $\Delta$ the cubic and higher terms in the potential become increasingly irrelevant.  So our quadratic approximation to the Schrodinger equation for the bulk wavefunctions in section \ref{sec:GeneralDecoupling} is justified.

In the introduction and in section \ref{sec:GeneralDecoupling} we mentioned that our methods can relate the density of states to the function $f(\Delta)$ that suppresses the contribution of dimension $\Delta$ operators to low-mass mesons and glueballs.  For the potential in equation (\ref{eq:ZAlphaPotential}) this is easy to see.  The energy of the $n$th level in the limit of large $n$ grows as the power-law
\be
m^2(n) \approx n^q \ \ \ \mathrm{with} \ \ \  q = {\frac{2 \alpha}{2 + \alpha}}
\ee
It's also true that the energy of the $n=0$ level grows at large $\Delta$ as $m^2(\Delta) \approx \Delta^q$.
The suppression factor
\be
f(\Delta) \approx \exp \left[-  \frac{\Delta^2 z_\Delta^2}{\Delta^2 + z_\Delta^4} \right] \propto e^{-  c \Delta^p} \ \ \ \mathrm{with} \ \ \  p = \min \left( q, {\frac{4}{2 + \alpha}} \right)
\ee
Thus we observe a simple relation that either $p = q$ for $\alpha < 2$ or $p = \frac{\alpha}{2} q$ for $\alpha > 2$ that follows because both parameters are derived from the same Schrodinger potential.

\section{Normalizations for Operators and Mesons}
\label{app:Normalizations}

CFT operators are usually normalized so that their engineering dimension equals their scaling dimension.  However, when conformal invariance is broken, it is more useful to normalize CFT operators using the LSZ prescription, so that operators have a fixed probability for creating massive particles.   The two normalizations differ by powers of dimensionful parameters, which we are able to determine in this appendix by comparing with hard-wall models in AdS/CFT.  Once the normalizations are fixed, we can make a physical comparison of the magnitude of 2-pt correlators between different operators.

\subsection{Normalizations for CFT Operators}
\label{sec:normalization}

We would like to motivate our study of decoupling as the general statement that two-point functions in broken CFTs behave like
\be
\label{eq:OpOverlapAppendix}
\< \CO_1(r) \CO_2(0) \> \sim \frac{f(\Delta_1, \Delta_2) e^{-m r}}{r^{d-2}}
\ee
where $f(\Delta_1, \Delta_2)$ is small when the dimension $\Delta_2$ is taken to be large.  But this statement requires some care.  The operators $\CO_1$ and $\CO_2$ are usually normalized so that they have engineering dimension equal to their scaling dimensions $\Delta_1$ and $\Delta_2$,  but in order for equation (\ref{eq:OpOverlapAppendix}) to make sense we must use a different normalization, where $\CO_1$ and $\CO_2$ have fixed engineering dimensions.  To compensate we must re-normalize $\CO_1$ and $\CO_2$ with a dimensionful coefficient.  

We will determine this coefficient by comparing with the results from AdS/CFT hard-wall models.  The momentum-space correlator can be written in a mode decomposition
\be
\< \CO(p) \CO(-p) \> &=& 
 \sum_n    \frac{ f_{n}^2  }{p^2 - m_n^2} 
\label{eq:cfttwopoints}
\ee
for the operator $\CO$ of dimension $\Delta$.  We can compare this to the behavior of a field $\Phi(x, z)$ in a slice of AdS to determine a standard value for $f_n$.  We can then renormalize $\CO \to \frac{1}{f_0} \CO$ if we wish in order to accord with the standard LSZ normalization for the lowest-lying mode.  

Recall the standard connection\footnote{See, for example \cite{Susskind:1998dq, Banks:1998dd, Harlow:2011ke}, the numeric normalization factor can be found in \cite{JoaoMellin}.}  between bulk operators $\Phi$ and the corresponding primary operators $\CO$ in the CFT:
\be
\CC_{\Delta}^{-1/2} \lim_{z \rightarrow 0} z^{-\Delta}  \Phi(z, x) &=& \CO(x), \ \ \mathrm{where} \ \ \CC_\Delta = \frac{\Gamma(\Delta)}{2\pi^{\frac{d}{2}}\Gamma(\Delta-\frac{d}{2} +1)}
\label{eq:bulkboundaryoperatorcorrespondence}
\ee
when $\CO(x)$ has been normalized to have a two-point function $\< \CO(x) \CO(0)\> = x^{-2\Delta}$.
We can take the Fourier transform of both sides of {\ref{eq:bulkboundaryoperatorcorrespondence} to relate $\Phi(z,p)$ to $\CO(p)$.  Now, consider a massive scalar field in an RS model with an IR boundary brane but no UV boundary brane.  The field $\Phi(z, p)$ has the mode decomposition
\be
\Phi(z,p) &=& \sum_n \psi_n(z) \phi_n(p).
\ee
This has the solution
\be
\psi_0(z) &=& N \sqrt{z} (kz)^{3/2} J_{\Delta-2}(m_0 z)
\ee
where $\Delta$ is the dimension of the CFT operator $\CO$ and $N$ is a normalization constant that is determined by
\be
\int_0^{z_{\rm IR}} dz \psi_0^2(z) = \frac{1}{k}.
\ee
Applying the correspondence (\ref{eq:bulkboundaryoperatorcorrespondence}), we easily extract that
\be
\CO(p) &=& \phi_0(p) \frac{N \CC_\Delta^{-1/2}}{\Gamma(\Delta-1)} k^{3/2}  \left( \frac{m_0}{2} \right)^{\Delta-2}+ \dots,
\ee
where $\dots$ are the higher KK modes.  By comparison with the two-point functions above (\ref{eq:cfttwopoints}), and using the fact that $\phi(p)$ is a canonically normalized 4d mode, we have obtained our expression for $f_{0}$ in the limit of ``order 1'' overlap:
\be
\label{eq:referencef}
f_{0} =  \frac{N\CC_\Delta^{-1/2}}{\Gamma(\Delta-1)} k^{3/2}  \left( \frac{m_0}{2} \right)^{\Delta-2}.
\ee
In general, $N$ is a moderately complicated function (it can be written down in closed form in terms of Bessel functions, but it is not particularly enlightening), but it simplifies significantly in the limit that $\Delta$ is large. In this limit, we expect that $m_0 \approx c \Delta/z_{IR}$ with $c$ of order 1, and at large $\Delta$ we find numerically that
\be
N \stackrel{\Delta \gg 1}{\approx} \frac{\Delta}{\sqrt{k} z_{\rm IR}} f(c)
\ee
Thus, we find that in the absence of any decoupling, we have overlap coefficients $f_{0}$ that behave parametrically as
\be
f_{0} &\approx& \left( \frac{m_0}{2}\right)^\Delta \frac{\CC_\Delta^{-1/2}}{\Gamma(\Delta-2)} \frac{4 k}{z_{\rm IR} m_0^2} f(c)
\ee
When we speak of small overlap coefficients, we therefore mean small compared to this reference value.

\subsection{Relationship with AdS Normalizations and Meson States}
\label{sec:normalization2}

In the body of the paper, we worked out the overlap of a heavy fermion in a slice of AdS with mixing terms on the IR boundary brane.  Here, we will extend the calculation to the case of scalars, in order to facilitate comparison with the overlap coefficients $f_{0}$ in the previous subsection.  Our action is simply two scalars $\Phi_{1,2}$ with bulk mass-squareds $m_1^2, m_2^2$ in pure AdS, plus a boundary mass mixing term:
\be
S \supset \int_{z=z_{\rm IR}} d^4 x \sqrt{g} \mu_{ab}^2 \Phi_a \Phi_b
\ee
We are interested in the bulk profile of the lightest KK mode, which sits inside both the $\Phi_1$ and $\Phi_2$ fields:
\be
\Phi_a(z,p) &= & \phi_0(p) g_{0a}(z) (k z)^{3/2} + \dots.
\label{eq:phidecomp}
\ee
Since the bulk is pure AdS, we can immediately write down the wavefunction for the lightest KK mode:
\be
g_{0a}(z) &=& \frac{y_a}{{\cal N}^{\frac{1}{2}}} \sqrt{z} J_{\Delta_a - 2}(m_0 z).
\ee
where the mass eigenvalue $m_0$ and the coefficients $y_a$ depend on the boundary conditions.  Generically, in a hard wall model, taking $\Delta_2$ to be large will cause $y_2$ to be small, but to keep things parallel with the fermionic example, we imagine tuning the boundary conditions so that $y_a$ and $m_0$ are kept constant as we increase $\Delta_2$.  As before, even in this tuned case, $\Phi_2$ decouples from the zero mode $\phi_0(p)$ at large $\Delta_2$.  As we take $\Delta_2$ with $m_0$ and $z_{\rm IR}$ held fixed, the Bessel function becomes approximated everywhere by its small $z$ behavior:
\be
g_{02}(z) &\approx& \frac{y_2}{{\cal N}^{\frac{1}{2}}} z^{\Delta_2 - \frac{3}{2}} \frac{1}{\Gamma(\Delta_2 -1)} \left( \frac{m_0}{2} \right)^{\Delta_2 -2}.
\label{eq:bulklargeDelta}
\ee
At large $\Delta_2$, this wave function falls off very rapidly at small $z$ and thus gives a small contribution to the normalization ${\cal N}$.  The normalization will therefore be determined mainly by the wavefunction $g_{01}$ component.  Since this is independent of $\Delta_2$, it is sufficient to simply parameterize it.  By inspection the normalization condition is ${\cal N}^{\frac{1}{2}} = b  \sqrt{k} z_{\rm IR}$ up to a numerical constant $b$ that does not concern us. We therefore have
\be
f_{\CO_2, 0} = \left( \frac{m_0}{2} \right)^{\Delta_2} \frac{4 y_2 k  }{b m_0^2 z_{\rm IR}} \frac{ \CC_{\Delta_2}^{-\frac{1}{2}}}{  \Gamma(\Delta_2 -1) }
\ee
which is parametrically down by a factor of $\Delta_2$ from the reference value (\ref{eq:referencef}).

In fact, it is straightforward to see generally why suppression in the overlap with the bulk field $g_{02}(z)$ will show up as a suppression in the overlap coefficient with the boundary operator, and vice versa.  In the limit of large $\Delta_2$ in a hard wall model, the bulk wavefunction reduces to a simple power-law (\ref{eq:bulklargeDelta}), which is exactly the $z$-dependence that is stripped off by the relation between the bulk field and the boundary CFT operator.  At large $\Delta_2$, there are no other features present in the wavefunction beyond the near-boundary behavior, so it is completely determined by the coefficient $f_{\CO_2, 0}$. Putting together equations (\ref{eq:phidecomp}) and (\ref{eq:bulkboundaryoperatorcorrespondence}), we see that at small $z$,
\be
\CO_2(p)   \approx \phi_0(p)\CC_\Delta^{-1/2}g_{0a}(z) k^{\frac{3}{2}} z^{\frac{3}{2}-\Delta_2} .
\ee
Taking the two-point function of $\CO(p)$ in this expression and using the fact that $\phi_0(p)$ is a canonically normalized field, so that $\< \phi_0(p) \phi_0(-p) \> = \frac{1}{p^2 - m_n^2}$, we can read off that at small $z$,  $g_{02}(z)$ must satisfy overlap coefficient $f_{\CO_2, 0}$:
\be
g_{02}(z) &\approx& f_{\CO_2, 0}  z^{\Delta_2 - \frac{3}{2}} \left( k^{-\frac{3}{2}} \CC_{\Delta_2}^{\frac{1}{2}} \right).
\ee
This relation is satisfied at small $z$ for any $\Delta_2$, large or small.  The fact that when $\Delta_2$ is large, $g_{02}(z)$ is everywhere approximated by its small $z$ behavior means that for large $\Delta_2$, $g_{02}(z)$ is completely determined by $f_{\CO_2, 0}$ and $\Delta_2$ through the above relation.  Unsuppressed overlap of the mode $\phi_0(p)$ with $\Phi_2$ means that it sits mostly inside $\Phi_2$, so the dominant contribution to its normalization comes from the $g_{02}$ component.  As a cross-check, we can see  that $g_{02}$ has approximately unit norm:
\be
1 &=& \sum_{i=1}^2 \int_0^{z_{\rm IR}} \frac{dz}{z} \frac{1}{(kz)^2} |g_{0i}(z)|^2 \approx \int_0^{z_{\rm IR}} \frac{dz}{z} \frac{1}{(kz)^2} |g_{02}(z)|^2 .
\ee
But since $g_{02}(z)$ is completely determined by $f_{\CO_2, 0}$, this normalization condition clearly sets $f_{\CO_2, 0}$ to be approximately its reference ``order 1'' value $f_0$ in (\ref{eq:referencef}).  Conversely, $f_{\CO_2, 0}$ much smaller than $f_0$ in (\ref{eq:referencef}) immediately implies that the contribution of $\int_0^{z_{\rm IR}} \frac{dz}{z} \frac{1}{(kz)^2} |g_{02}(z)|^2$ to the full norm of $g_{0a}(z)$ is much less than 1.  Thus, for large dimension operators, small overlap with a KK mode can be equivalently diagnosed through the size of the overlap of the KK mode wavefunction $g_{02}(z)$ in AdS or through the size of the ratio of the overlap coefficient $f_{\CO_2,0}$ to $f_0$.

\bibliographystyle{utphys}
\bibliography{EFTandBrokenCFT}

\providecommand{\href}[2]{#2}\begingroup\raggedright\begin{thebibliography}{10}

\bibitem{Maldacena:1997re}
J.~M. Maldacena, ``{The Large N limit of superconformal field theories and
  supergravity},'' \href{http://dx.doi.org/10.1023/A:1026654312961,
  10.1023/A:1026654312961}{{\em Adv.Theor.Math.Phys.} {\bfseries 2} (1998)
  231--252}, \href{http://arxiv.org/abs/hep-th/9711200}{{\ttfamily
  arXiv:hep-th/9711200 [hep-th]}}.

\bibitem{Witten}
E.~Witten, ``{Anti-de Sitter space and holography},'' {\em Adv. Theor. Math.
  Phys.} {\bfseries 2} (1998) 253--291,
\href{http://arxiv.org/abs/hep-th/9802150}{{\ttfamily arXiv:hep-th/9802150}}.
%%CITATION = HEP-TH/9802150;%%.

\bibitem{Gubser:1998bc}
S.~Gubser, I.~R. Klebanov, and A.~M. Polyakov, ``{Gauge theory correlators from
  noncritical string theory},''
  \href{http://dx.doi.org/10.1016/S0370-2693(98)00377-3}{{\em Phys.Lett.}
  {\bfseries B428} (1998) 105--114},
  \href{http://arxiv.org/abs/hep-th/9802109}{{\ttfamily arXiv:hep-th/9802109
  [hep-th]}}.

\bibitem{Randall:1999ee}
L.~Randall and R.~Sundrum, ``{A Large mass hierarchy from a small extra
  dimension},'' \href{http://dx.doi.org/10.1103/PhysRevLett.83.3370}{{\em
  Phys.Rev.Lett.} {\bfseries 83} (1999) 3370--3373},
\href{http://arxiv.org/abs/hep-ph/9905221}{{\ttfamily arXiv:hep-ph/9905221
  [hep-ph]}}.
%%CITATION = HEP-PH/9905221;%%.

\bibitem{Randall:1999vf}
L.~Randall and R.~Sundrum, ``{An Alternative to compactification},''
  \href{http://dx.doi.org/10.1103/PhysRevLett.83.4690}{{\em Phys.Rev.Lett.}
  {\bfseries 83} (1999) 4690--4693},
\href{http://arxiv.org/abs/hep-th/9906064}{{\ttfamily arXiv:hep-th/9906064
  [hep-th]}}.
%%CITATION = HEP-TH/9906064;%%.

\bibitem{Karch:2002sh}
A.~Karch and E.~Katz, ``{Adding flavor to AdS / CFT},'' {\em JHEP} {\bfseries
  0206} (2002) 043,
\href{http://arxiv.org/abs/hep-th/0205236}{{\ttfamily arXiv:hep-th/0205236
  [hep-th]}}.
%%CITATION = HEP-TH/0205236;%%.

\bibitem{Erlich:2005qh}
J.~Erlich, E.~Katz, D.~T. Son, and M.~A. Stephanov, ``{QCD and a holographic
  model of hadrons},''
  \href{http://dx.doi.org/10.1103/PhysRevLett.95.261602}{{\em Phys.Rev.Lett.}
  {\bfseries 95} (2005) 261602},
\href{http://arxiv.org/abs/hep-ph/0501128}{{\ttfamily arXiv:hep-ph/0501128
  [hep-ph]}}.
%%CITATION = HEP-PH/0501128;%%.

\bibitem{Karch:2006pv}
A.~Karch, E.~Katz, D.~T. Son, and M.~A. Stephanov, ``{Linear confinement and
  AdS/QCD},'' \href{http://dx.doi.org/10.1103/PhysRevD.74.015005}{{\em
  Phys.Rev.} {\bfseries D74} (2006) 015005},
\href{http://arxiv.org/abs/hep-ph/0602229}{{\ttfamily arXiv:hep-ph/0602229
  [hep-ph]}}.
%%CITATION = HEP-PH/0602229;%%.

\bibitem{Hartnoll:2008vx}
S.~A. Hartnoll, C.~P. Herzog, and G.~T. Horowitz, ``{Building a Holographic
  Superconductor},''
  \href{http://dx.doi.org/10.1103/PhysRevLett.101.031601}{{\em Phys.Rev.Lett.}
  {\bfseries 101} (2008) 031601},
\href{http://arxiv.org/abs/0803.3295}{{\ttfamily arXiv:0803.3295 [hep-th]}}.
%%CITATION = ARXIV:0803.3295;%%.

\bibitem{Balasubramanian:2008dm}
K.~Balasubramanian and J.~McGreevy, ``{Gravity duals for non-relativistic
  CFTs},'' \href{http://dx.doi.org/10.1103/PhysRevLett.101.061601}{{\em
  Phys.Rev.Lett.} {\bfseries 101} (2008) 061601},
\href{http://arxiv.org/abs/0804.4053}{{\ttfamily arXiv:0804.4053 [hep-th]}}.
%%CITATION = ARXIV:0804.4053;%%.

\bibitem{Son:2008ye}
D.~Son, ``{Toward an AdS/cold atoms correspondence: A Geometric realization of
  the Schrodinger symmetry},''
  \href{http://dx.doi.org/10.1103/PhysRevD.78.046003}{{\em Phys.Rev.}
  {\bfseries D78} (2008) 046003},
\href{http://arxiv.org/abs/0804.3972}{{\ttfamily arXiv:0804.3972 [hep-th]}}.
%%CITATION = ARXIV:0804.3972;%%.

\bibitem{Liu:2009dm}
H.~Liu, J.~McGreevy, and D.~Vegh, ``{Non-Fermi liquids from holography},''
  \href{http://dx.doi.org/10.1103/PhysRevD.83.065029}{{\em Phys.Rev.}
  {\bfseries D83} (2011) 065029},
\href{http://arxiv.org/abs/0903.2477}{{\ttfamily arXiv:0903.2477 [hep-th]}}.
%%CITATION = ARXIV:0903.2477;%%.

\bibitem{JP}
I.~Heemskerk, J.~Penedones, J.~Polchinski, and J.~Sully, ``{Holography from
  Conformal Field Theory},''
  \href{http://dx.doi.org/10.1088/1126-6708/2009/10/079}{{\em JHEP} {\bfseries
  10} (2009) 079},
\href{http://arxiv.org/abs/0907.0151}{{\ttfamily arXiv:0907.0151 [hep-th]}}.
%%CITATION = 0907.0151;%%.

\bibitem{Katz}
A.~L. Fitzpatrick, E.~Katz, D.~Poland, and D.~Simmons-Duffin, ``{Effective
  Conformal Theory and the Flat-Space Limit of AdS},''
  \href{http://dx.doi.org/10.1007/JHEP07(2011)023}{{\em JHEP} {\bfseries 1107}
  (2011) 023},
\href{http://arxiv.org/abs/1007.2412}{{\ttfamily arXiv:1007.2412 [hep-th]}}.
%%CITATION = ARXIV:1007.2412;%%.

\bibitem{Fitzpatrick:2011hh}
A.~L. Fitzpatrick and D.~Shih, ``{Anomalous Dimensions of Non-Chiral Operators
  from AdS/CFT},'' \href{http://dx.doi.org/10.1007/JHEP10(2011)113}{{\em JHEP}
  {\bfseries 10} (2011) 113},
\href{http://arxiv.org/abs/1104.5013}{{\ttfamily arXiv:1104.5013 [hep-th]}}.
%%CITATION = 1104.5013;%%.

\bibitem{NaturalLanguage}
A.~L. Fitzpatrick, J.~Kaplan, J.~Penedones, S.~Raju, and B.~C. van Rees, ``{A
  Natural Language for AdS/CFT Correlators},''
  \href{http://dx.doi.org/10.1007/JHEP11(2011)095}{{\em JHEP} {\bfseries 1111}
  (2011) 095},
\href{http://arxiv.org/abs/1107.1499}{{\ttfamily arXiv:1107.1499 [hep-th]}}.
%%CITATION = ARXIV:1107.1499;%%.

\bibitem{ElShowk:2011ag}
S.~El-Showk and K.~Papadodimas, ``{Emergent Spacetime and Holographic CFTs},''
  \href{http://dx.doi.org/10.1007/JHEP10(2012)106}{{\em JHEP} {\bfseries 1210}
  (2012) 106},
\href{http://arxiv.org/abs/1101.4163}{{\ttfamily arXiv:1101.4163 [hep-th]}}.
%%CITATION = ARXIV:1101.4163;%%.

\bibitem{AdSfromCFT}
A.~L. Fitzpatrick and J.~Kaplan, ``{AdS Field Theory from Conformal Field
  Theory},''
\href{http://arxiv.org/abs/1208.0337}{{\ttfamily arXiv:1208.0337 [hep-th]}}.
%%CITATION = ARXIV:1208.0337;%%.

\bibitem{Heemskerk:2010ty}
I.~Heemskerk and J.~Sully, ``{More Holography from Conformal Field Theory},''
  \href{http://dx.doi.org/10.1007/JHEP09(2010)099}{{\em JHEP} {\bfseries 1009}
  (2010) 099},
\href{http://arxiv.org/abs/1006.0976}{{\ttfamily arXiv:1006.0976 [hep-th]}}.
%%CITATION = ARXIV:1006.0976;%%.

\bibitem{Sundrum:2011ic}
R.~Sundrum, ``{From Fixed Points to the Fifth Dimension},''
  \href{http://dx.doi.org/10.1103/PhysRevD.86.085025}{{\em Phys.Rev.}
  {\bfseries D86} (2012) 085025},
\href{http://arxiv.org/abs/1106.4501}{{\ttfamily arXiv:1106.4501 [hep-th]}}.
%%CITATION = ARXIV:1106.4501;%%.

\bibitem{Fubini:1972mf}
S.~Fubini, A.~J. Hanson, and R.~Jackiw, ``{New approach to field theory},''
\href{http://dx.doi.org/10.1103/PhysRevD.7.1732}{{\em Phys.Rev.} {\bfseries D7}
  (1973) 1732--1760}.
%%CITATION = PHRVA,D7,1732;%%.

\bibitem{tHooft:1974hx}
G.~'t~Hooft, ``{A Two-Dimensional Model for Mesons},''
\href{http://dx.doi.org/10.1016/0550-3213(74)90088-1}{{\em Nucl.Phys.}
  {\bfseries B75} (1974) 461}.
%%CITATION = NUPHA,B75,461;%%.

\bibitem{Katz:2007br}
E.~Katz and T.~Okui, ``{The 't Hooft model as a hologram},''
  \href{http://dx.doi.org/10.1088/1126-6708/2009/01/013}{{\em JHEP} {\bfseries
  0901} (2009) 013},
\href{http://arxiv.org/abs/0710.3402}{{\ttfamily arXiv:0710.3402 [hep-th]}}.
%%CITATION = ARXIV:0710.3402;%%.

\bibitem{Freedman:1999gp}
D.~Freedman, S.~Gubser, K.~Pilch, and N.~Warner, ``{Renormalization group flows
  from holography supersymmetry and a c theorem},'' {\em Adv.Theor.Math.Phys.}
  {\bfseries 3} (1999) 363--417,
\href{http://arxiv.org/abs/hep-th/9904017}{{\ttfamily arXiv:hep-th/9904017
  [hep-th]}}.
%%CITATION = HEP-TH/9904017;%%.

\bibitem{Girardello:1998pd}
L.~Girardello, M.~Petrini, M.~Porrati, and A.~Zaffaroni, ``{Novel local CFT and
  exact results on perturbations of N=4 superYang Mills from AdS dynamics},''
  {\em JHEP} {\bfseries 9812} (1998) 022,
\href{http://arxiv.org/abs/hep-th/9810126}{{\ttfamily arXiv:hep-th/9810126
  [hep-th]}}.
%%CITATION = HEP-TH/9810126;%%.

\bibitem{Myers:2012ed}
R.~C. Myers and A.~Singh, ``{Comments on Holographic Entanglement Entropy and
  RG Flows},'' \href{http://dx.doi.org/10.1007/JHEP04(2012)122}{{\em JHEP}
  {\bfseries 1204} (2012) 122},
\href{http://arxiv.org/abs/1202.2068}{{\ttfamily arXiv:1202.2068 [hep-th]}}.
%%CITATION = ARXIV:1202.2068;%%.

\bibitem{Gherghetta:2006ha}
T.~Gherghetta, ``{Les Houches lectures on warped models and holography},''
\href{http://arxiv.org/abs/hep-ph/0601213}{{\ttfamily arXiv:hep-ph/0601213
  [hep-ph]}}.
%%CITATION = HEP-PH/0601213;%%.

\bibitem{GoldbergerWise}
W.~D. Goldberger and M.~B. Wise, ``{Modulus stabilization with bulk fields},''
  \href{http://dx.doi.org/10.1103/PhysRevLett.83.4922}{{\em Phys.Rev.Lett.}
  {\bfseries 83} (1999) 4922--4925},
\href{http://arxiv.org/abs/hep-ph/9907447}{{\ttfamily arXiv:hep-ph/9907447
  [hep-ph]}}.
%%CITATION = HEP-PH/9907447;%%.

\bibitem{Georgi:2000ks}
H.~Georgi, A.~K. Grant, and G.~Hailu, ``{Brane couplings from bulk loops},''
  \href{http://dx.doi.org/10.1016/S0370-2693(01)00408-7}{{\em Phys.Lett.}
  {\bfseries B506} (2001) 207--214},
\href{http://arxiv.org/abs/hep-ph/0012379}{{\ttfamily arXiv:hep-ph/0012379
  [hep-ph]}}.
%%CITATION = HEP-PH/0012379;%%.

\bibitem{Cheng:2002iz}
H.-C. Cheng, K.~T. Matchev, and M.~Schmaltz, ``{Radiative corrections to
  Kaluza-Klein masses},''
  \href{http://dx.doi.org/10.1103/PhysRevD.66.036005}{{\em Phys.Rev.}
  {\bfseries D66} (2002) 036005},
\href{http://arxiv.org/abs/hep-ph/0204342}{{\ttfamily arXiv:hep-ph/0204342
  [hep-ph]}}.
%%CITATION = HEP-PH/0204342;%%.

\bibitem{Delaunay:2012cz}
C.~Delaunay, J.~F. Kamenik, G.~Perez, and L.~Randall, ``{Charming CP Violation
  and Dipole Operators from RS Flavor Anarchy},''
  \href{http://dx.doi.org/10.1007/JHEP01(2013)027}{{\em JHEP} {\bfseries 1301}
  (2013) 027},
\href{http://arxiv.org/abs/1207.0474}{{\ttfamily arXiv:1207.0474 [hep-ph]}}.
%%CITATION = ARXIV:1207.0474;%%.

\bibitem{Grossman:1999ra}
Y.~Grossman and M.~Neubert, ``{Neutrino masses and mixings in nonfactorizable
  geometry},'' \href{http://dx.doi.org/10.1016/S0370-2693(00)00054-X}{{\em
  Phys.Lett.} {\bfseries B474} (2000) 361--371},
\href{http://arxiv.org/abs/hep-ph/9912408}{{\ttfamily arXiv:hep-ph/9912408
  [hep-ph]}}.
%%CITATION = HEP-PH/9912408;%%.

\bibitem{ArkaniHamed:2001is}
N.~Arkani-Hamed, A.~G. Cohen, and H.~Georgi, ``{Anomalies on orbifolds},''
  \href{http://dx.doi.org/10.1016/S0370-2693(01)00946-7}{{\em Phys.Lett.}
  {\bfseries B516} (2001) 395--402},
\href{http://arxiv.org/abs/hep-th/0103135}{{\ttfamily arXiv:hep-th/0103135
  [hep-th]}}.
%%CITATION = HEP-TH/0103135;%%.

\bibitem{Hanson}
T.~Eguchi, P.~B. Gilkey, and A.~J. Hanson, ``{Gravitation, Gauge Theories and
  Differential Geometry},''
\href{http://dx.doi.org/10.1016/0370-1573(80)90130-1}{{\em Phys.Rept.}
  {\bfseries 66} (1980) 213}.
%%CITATION = PRPLC,66,213;%%.

\bibitem{Zamolodchikov:1986gt}
A.~Zamolodchikov, ``{Irreversibility of the Flux of the Renormalization Group
  in a 2D Field Theory},''
{\em JETP Lett.} {\bfseries 43} (1986) 730--732.
%%CITATION = JTPLA,43,730;%%.

\bibitem{Komargodski:2011vj}
Z.~Komargodski and A.~Schwimmer, ``{On Renormalization Group Flows in Four
  Dimensions},'' \href{http://dx.doi.org/10.1007/JHEP12(2011)099}{{\em JHEP}
  {\bfseries 1112} (2011) 099},
\href{http://arxiv.org/abs/1107.3987}{{\ttfamily arXiv:1107.3987 [hep-th]}}.
%%CITATION = ARXIV:1107.3987;%%.

\bibitem{Komargodski:2011xv}
Z.~Komargodski, ``{The Constraints of Conformal Symmetry on RG Flows},''
  \href{http://dx.doi.org/10.1007/JHEP07(2012)069}{{\em JHEP} {\bfseries 1207}
  (2012) 069},
\href{http://arxiv.org/abs/1112.4538}{{\ttfamily arXiv:1112.4538 [hep-th]}}.
%%CITATION = ARXIV:1112.4538;%%.

\bibitem{Klebanov:2000hb}
I.~R. Klebanov and M.~J. Strassler, ``{Supergravity and a confining gauge
  theory: Duality cascades and chi SB resolution of naked singularities},''
  {\em JHEP} {\bfseries 0008} (2000) 052,
\href{http://arxiv.org/abs/hep-th/0007191}{{\ttfamily arXiv:hep-th/0007191
  [hep-th]}}.
%%CITATION = HEP-TH/0007191;%%.

\bibitem{Polchinski:2000uf}
J.~Polchinski and M.~J. Strassler, ``{The String dual of a confining
  four-dimensional gauge theory},''
\href{http://arxiv.org/abs/hep-th/0003136}{{\ttfamily arXiv:hep-th/0003136
  [hep-th]}}.
%%CITATION = HEP-TH/0003136;%%.

\bibitem{Kruczenski:2003be}
M.~Kruczenski, D.~Mateos, R.~C. Myers, and D.~J. Winters, ``{Meson spectroscopy
  in AdS / CFT with flavor},'' {\em JHEP} {\bfseries 0307} (2003) 049,
\href{http://arxiv.org/abs/hep-th/0304032}{{\ttfamily arXiv:hep-th/0304032
  [hep-th]}}.
%%CITATION = HEP-TH/0304032;%%.

\bibitem{Sakai:2004cn}
T.~Sakai and S.~Sugimoto, ``{Low energy hadron physics in holographic QCD},''
  \href{http://dx.doi.org/10.1143/PTP.113.843}{{\em Prog.Theor.Phys.}
  {\bfseries 113} (2005) 843--882},
\href{http://arxiv.org/abs/hep-th/0412141}{{\ttfamily arXiv:hep-th/0412141
  [hep-th]}}.
%%CITATION = HEP-TH/0412141;%%.

\bibitem{Sakai:2005yt}
T.~Sakai and S.~Sugimoto, ``{More on a holographic dual of QCD},''
  \href{http://dx.doi.org/10.1143/PTP.114.1083}{{\em Prog.Theor.Phys.}
  {\bfseries 114} (2005) 1083--1118},
\href{http://arxiv.org/abs/hep-th/0507073}{{\ttfamily arXiv:hep-th/0507073
  [hep-th]}}.
%%CITATION = HEP-TH/0507073;%%.

\bibitem{Batell:2008zm}
B.~Batell and T.~Gherghetta, ``{Dynamical Soft-Wall AdS/QCD},''
  \href{http://dx.doi.org/10.1103/PhysRevD.78.026002}{{\em Phys.Rev.}
  {\bfseries D78} (2008) 026002},
\href{http://arxiv.org/abs/0801.4383}{{\ttfamily arXiv:0801.4383 [hep-ph]}}.
%%CITATION = ARXIV:0801.4383;%%.

\bibitem{Myers:2010tj}
R.~C. Myers and A.~Sinha, ``{Holographic c-theorems in arbitrary dimensions},''
  \href{http://dx.doi.org/10.1007/JHEP01(2011)125}{{\em JHEP} {\bfseries 1101}
  (2011) 125},
\href{http://arxiv.org/abs/1011.5819}{{\ttfamily arXiv:1011.5819 [hep-th]}}.
%%CITATION = ARXIV:1011.5819;%%.

\bibitem{Karch:2010eg}
A.~Karch, E.~Katz, D.~T. Son, and M.~A. Stephanov, ``{On the sign of the
  dilaton in the soft wall models},''
  \href{http://dx.doi.org/10.1007/JHEP04(2011)066}{{\em JHEP} {\bfseries 1104}
  (2011) 066},
\href{http://arxiv.org/abs/1012.4813}{{\ttfamily arXiv:1012.4813 [hep-ph]}}.
%%CITATION = ARXIV:1012.4813;%%.

\bibitem{Hong:2003jm}
S.~Hong, S.~Yoon, and M.~J. Strassler, ``{Quarkonium from the
  fifth-dimension},''
  \href{http://dx.doi.org/10.1088/1126-6708/2004/04/046}{{\em JHEP} {\bfseries
  0404} (2004) 046},
\href{http://arxiv.org/abs/hep-th/0312071}{{\ttfamily arXiv:hep-th/0312071
  [hep-th]}}.
%%CITATION = HEP-TH/0312071;%%.

\bibitem{Hong:2004gz}
S.~Hong, S.~Yoon, and M.~J. Strassler, ``{Adjoint trapping: A New phenomenon at
  strong 't Hooft coupling},''
  \href{http://dx.doi.org/10.1088/1126-6708/2006/03/012}{{\em JHEP} {\bfseries
  0603} (2006) 012},
\href{http://arxiv.org/abs/hep-th/0410080}{{\ttfamily arXiv:hep-th/0410080
  [hep-th]}}.
%%CITATION = HEP-TH/0410080;%%.

\bibitem{Csaki:2008dt}
C.~Csaki, M.~Reece, and J.~Terning, ``{The AdS/QCD Correspondence: Still
  Undelivered},'' \href{http://dx.doi.org/10.1088/1126-6708/2009/05/067}{{\em
  JHEP} {\bfseries 0905} (2009) 067},
\href{http://arxiv.org/abs/0811.3001}{{\ttfamily arXiv:0811.3001 [hep-ph]}}.
%%CITATION = ARXIV:0811.3001;%%.

\bibitem{Susskind:1998dq}
L.~Susskind and E.~Witten, ``{The Holographic bound in anti-de Sitter space},''
\href{http://arxiv.org/abs/hep-th/9805114}{{\ttfamily arXiv:hep-th/9805114
  [hep-th]}}.
%%CITATION = HEP-TH/9805114;%%.

\bibitem{Banks:1998dd}
T.~Banks, M.~R. Douglas, G.~T. Horowitz, and E.~J. Martinec, ``{AdS dynamics
  from conformal field theory},''
  \href{http://arxiv.org/abs/hep-th/9808016}{{\ttfamily arXiv:hep-th/9808016
  [hep-th]}}.

\bibitem{Harlow:2011ke}
D.~Harlow and D.~Stanford, ``{Operator Dictionaries and Wave Functions in
  AdS/CFT and dS/CFT},'' \href{http://arxiv.org/abs/1104.2621}{{\ttfamily
  arXiv:1104.2621 [hep-th]}}.

\bibitem{JoaoMellin}
J.~Penedones, ``{Writing CFT correlation functions as AdS scattering
  amplitudes},'' \href{http://dx.doi.org/10.1007/JHEP03(2011)025}{{\em JHEP}
  {\bfseries 03} (2011) 025},
\href{http://arxiv.org/abs/1011.1485}{{\ttfamily arXiv:1011.1485 [hep-th]}}.
%%CITATION = 1011.1485;%%.

\end{thebibliography}\endgroup

\end{document}